**Thermodynamic characterization of the ($H_2$ + $C_3H_8$) system significant for the hydrogen economy: Experimental ($p$, $\rho$, $T$) determination and equation of state modelling**

Daniel Lozano-Martín[1], Peyman Khanipourmehrin[2], Heinrich Kipphardt[2], Dirk Tuma[2], and César R. Chamorro[1,*]

[1] Grupo de Termodinámica y Calibración (TERMOCAL), Research Institute on Bioeconomy, Escuela de Ingenierías Industriales, Universidad de Valladolid, Paseo del Cauce, 59, E-47011 Valladolid, Spain.

[2] BAM Bundesanstalt für Materialforschung und -prüfung, D-12200 Berlin, Germany.

## Abstract

For the gradual introduction of hydrogen in the energy market, the study of the properties of mixtures of hydrogen with typical components of natural gas (NG) and liquified petroleum gas (LPG) is of great importance. This work aims to provide accurate experimental ($p$, $\rho$, $T$) data for three hydrogen-propane mixtures with nominal compositions (amount of substance, mol/mol) of (0.95 $H_2$ + 0.05 $C_3H_8$), (0.90 $H_2$ + 0.10 $C_3H_8$), and (0.83 $H_2$ + 0.17 $C_3H_8$), at temperatures of 250, 275, 300, 325, 350, and 375 K, and pressures up to 20 MPa. A single-sinker densimeter was used to determine the density of the mixtures. Experimental density data were compared to the densities calculated from two reference equations of state: the GERG-2008 and the AGA8-DC92. Relative deviations from the GERG-2008 EoS are systematically larger than from the AGA8-DC92. They are within the ±0.5 % band, for the mixture with 5 % of propane, but deviations are higher than 0.5 % for the mixtures with 10 % and 17 % of propane, especially at low temperatures and high pressures. Finally, the sets of new experimental data have been processed by the application of two different statistical equations of state: the virial equation of state, through the second and third virial coefficients, $B(T, x)$ and $C(T, x)$, and the PC-SAFT equation of state.

* Corresponding author e-mail: cescha@eii.uva.es. Tel.: +34 983185697. Fax: +34 983423363

## 1. Introduction

Due to both urgent political and climate issues, the energy industry is rapidly moving towards a decarbonization of the system and one of the pillars of this change is hydrogen [1,2]. Whether used directly as a zero-$CO_2$ fuel or as energy carrier for the synthesis of alternative fuels [3], as long as the primary source of this hydrogen is the electrolysis of water using renewable energy sources (green hydrogen) or the reforming of fossil fuels with associated $CO_2$ capture processes (blue hydrogen) [4], the criterion of zero net carbon emissions is fulfilled [5]. However, the low energy density per unit volume of hydrogen compared to that of compressed natural gas or even petrol makes it necessary to store large quantities of produced hydrogen under high pressure [6].

One of the possibilities is to store hydrogen underground [7,8], for example, in depleted natural gas fields [9], deep aquifers [10] or salt caverns [11]. In these cases, a cushioning gas mixed with hydrogen is required to maintain the pressure inside the reservoir. Potential locations in various countries have been analyzed by several authors [12–16]. Some numerical case studies have covered the simulation of underground hydrogen injection and withdrawal in aquifers and salt caverns [17], where nitrogen or $CO_2$ are used as the cushion gas due to their availability and high compressibility [18]. The results of these simulations show that the most relevant drawbacks for a technically and economically feasible hydrogen storage are related to the high mobility and reactivity of hydrogen [19]. In the case of old natural gas reservoirs, the small residual amount of natural gas remaining can have the effect of a cushion gas. However, the number of investigations that examine this possibility is rather limited [20–23]. They agree that the acceptable volume of cushion gas with respect to the working gas should be between (20 to 80) vol-%. They show that the thermodynamic properties of hydrogen blended with the cushion gas, i.e., natural gas, is a key factor to optimize the injection-withdrawal cyclic process, maintain the purity of the hydrogen throughout the cycle and reduce hydrogen losses. Therefore, considering the scarcity of data available in the literature, it is important to study mixtures of hydrocarbons and other typical components of natural gas in mixtures highly enriched in hydrogen.

In addition, blending hydrogen with liquefied petroleum gas (LPG) is gaining more attention, especially for its use in internal combustion engines, as demonstrated in several studies [24–26]. LPG is defined as a mixture of (propane + butane) with a butane concentration between (20 up to 80) vol-%, with a typical composition of (0.70 $C_4H_{10}$ + 0.30 $C_3H_8$). LPG is obtained from the production of natural gas and petroleum refining and has been used as a fuel in several commercial engines since the last decade, as the exhaust gases are less polluting compared to those from the combustion of gasoline or diesel. In a similar way to how gasoline and diesel have become more sustainable by adding increasing proportions of renewable compounds, such as bioethanol or biodiesel, to the fuel that is currently in use [27]; hydrogen has been proposed as an additive for LPG during the transition period into a $CO_2$-free economy. A precise knowledge of the corresponding thermodynamic properties of hydrogen with the constituents of LPG is essential for the design and operation of all transport, storage and separation processes for these blends.

This work focuses on the experimental determination and the modeling for several equations of state of the volumetric properties of three hydrogen-propane mixtures with nominal compositions (amount of substance mol/mol) of (0.95 $H_2$ + 0.05 $C_3H_8$), (0.90 $H_2$ + 0.10 $C_3H_8$), and ultimately (0.83 $H_2$ + 0.17 $C_3H_8$).

One of the most accurate techniques over wide ranges of temperature and pressure has been used for the experimental part, namely a single-sinker densimeter with magnetic suspension coupling. This kind of apparatus has been used in other works for the determination of ($p$, $\rho$, $T$) data related to hydrogen-containing mixtures [28]: Jaeschke et al. [29] studied the systems ($CH_4$ + $H_2$), ($CO_2$ + $H_2$); and ($C_2H_6$ + $H_2$), Ben Souissi et al. [30] investigated the ($CO_2$ + $H_2$) mixture; Scholz and Span [31] studied the (Ar + $H_2$) mixture; Richter et al. [32] measured three multicomponent natural-gas samples blended with hydrogen; and Hernández-Gómez et al. [33–35] carried out a study on binary ($CH_4$ + $H_2$) and ($N_2$ + $H_2$) mixtures and a synthetic multicomponent $H_2$-enriched natural gas.

Regarding the most recent works aimed at determining the density for binary hydrogen blends, these focus on the study of the system ($CO_2$ + $H_2$). The thermodynamic characterization of this mixture is relevant in both carbon capture and storage technologies, as well as in the use of supercritical $CO_2$ as the solvation medium in hydrogenation reactions. The former is carried out in the research of Alsiyabi [36] and Sanchez-Vicente et al. [37] with a vibrating U-tube densimeter; and Tsankova et al. [38], with a re-entrant resonance cavity; while the latter is performed in the works of Cipollina et al. [39] and Zhang et al. [40] by means of the isochoric method.

The density of several multicomponent mixtures enriched with hydrogen has also been determined by Cheng et al. [41] for the system ($H_2$ + $CO_2$ and $H_2$ + $CO_2$ + $CH_4$) using the Burnett method and the system ($H_2$ + $CO_2$ + $CH_4$ + CO + $H_2O$) with the isochoric method. These mixtures are interesting for the development of a proposed compact gas turbine energy cycle for electricity and hydrogen production [42], from the gasification of fossil fuels in a supercritical water reactor with an implemented carbon capture and storage stage. The optimization of all the parts involved in the power generation cycle has also required several simulations of mixtures of hydrogen and supercritical water with the other gases present in the whole process [43,44], as well as the experimental determination of transport properties [45,46].

Further available data on density and compressibility factors for binary $H_2$ mixtures are reported elsewhere [47], describing the experimental techniques used, the composition, temperature, and pressure ranges, as well as the stated uncertainties in the different sources.

The experimental density results of this research have been compared to the two established reference Helmholtz-type multiparametric equations of state (EoS) for natural gas mixtures, namely the AGA8-DC92 EoS developed by the American Gas Association [48,49] and the GERG-2008 EoS proposed by the Groupe Européen de Recherches Gazières [50,51], respectively. 227 new experimental density results, for three different ($H_2$ + $C_3H_8$) mixtures, at six different temperatures from 250 K to 375 K, and at pressures up to 20 MPa, are presented in this work. The only volumetric data found in the literature for this binary system are in the works of Mihara et al. [52] and Mason and Eakin [53]. The data sets of Mihara et al. comprise 73 experimental points for 3 mixtures with a percentage molar fraction of propane between (16 to 27) mol-%, which is higher than those of this research, at pressures below 5 MPa and temperatures within the range from 298.15 to 348.15 K. The data set of Mason and Eakin -includes just 2 points at ambient pressure for two equimolar mixtures at 288.7 K. Mihara et al. [52] report a standard ($k$ = 1) uncertainty in density of 0.07 % for their measurements, determined by the Burnett method. These data were not used for the fitting procedure of the GERG-2008 EoS [50,51] but they were used for testing its validity range. The only other available ($p$, $\rho$, $T$) data which involve hydrogen and propane originate from the paper by Hegel et al. [54] which reports both experimental and modeled density data of hydrogen + propane + sunflower oil mixtures. These mixtures are of interest for the hydrogenation of this vegetable oil in a supercritical reactor.

Finally, the sets of new experimental data of this work have been processed by the application of two different statistical equations: the virial equation of state through the second and third virial coefficients, $B(T,x)$ and $C(T,x)$, as well as the second interaction virial coefficient $B_{12}(T)$, and the PC-SAFT EoS [55–57]. Both equations, widely used in engineering, are based on modeling the intermolecular forces between fluid molecules and, therefore, can give a physical interpretation at the microscopic level.

SAFT-type equations have already been used successfully in the modeling of systems of interest for $CO_2$ capture [58] and also for natural-gas-type mixtures [59,60]. However, those works in which they have been used to study mixtures with hydrogen are rather limited [61–63]. The work of Köster et al. [61] applies the PC-SAFT equation to hydrogen, nitrogen, oxygen, argon, water and their combinations; whereas in the works of Alkhatib et al. [62,63], the modeling is based on the polar soft-SAFT, rather than the most common PC-SAFT. In fact, our work is the first to apply the PC-SAFT EoS to mixtures of hydrogen with alkanes. In addition, works related to the assessment of the effect of hydrogen on the thermophysical properties of liquefied natural gas systems have required the development of quantum corrections for proper modeling with cubic equations of state [64] and Statistical Associating Fluid Theory for Mie potentials [65,66].

## 2. Experimental method

### 2.1. Mixture preparation

The three ($H_2$ + $C_3H_8$) mixtures investigated were prepared by the gravimetric procedure following the standard ISO 6142-1 [67] at the Federal Institute for Materials Research and Testing (Bundesanstalt für Materialforschung und -prüfung, BAM) in a similar way as to that outlined in detail elsewhere [68], using an electronic comparator balance (Sartorius LA 34000P-0CE, Sartorius AG) and a high-precision mechanical balance (Voland HCE 25, Voland Corp.) for the determination of the introduced mass portions. The certified gravimetric composition in molar percentage $x_i$ and the corresponding expanded ($k$ = 2) uncertainty in absolute terms $U(x_i)$ are reported in Table 1, together with the purity of the pure gases used for the three mixtures. Both hydrogen and propane were used without further purification, but the information on impurities by the supplier was considered for the specification of the uncertainty $U(x_i)$.

The preparation of a cylinder (aluminum alloy L6X®, $V$ = 10 dm$^3$, Luxfer Gas Cylinders) consisted of two consecutive filling steps. The calculated amount of propane is directly introduced from the storage cylinder into the recipient cylinder in the first step; the second step is the addition of hydrogen to finish the mixture. The mass portions and the final cylinder pressures are given below. After homogenization by subsequent heating and rolling for about 8 hours, the gas samples were ready for validation.

Cylinder no. 2020-200803 ($x_{H_2}$ = 0.95): 122.763 g of $C_3H_8$, 106.715 g of $H_2$, $p_{cylinder}$ = 14.7 MPa.

Cylinder no. 2023-200803 ($x_{H_2}$ = 0.90): 255.780 g of $C_3H_8$, 105.328 g of $H_2$, $p_{cylinder}$ = 14.8 MPa.

Cylinder no. 96054 989-210629 ($x_{H_2}$ = 0.83): 187.836 g of $C_3H_8$, 42.113 g of $H_2$, $p_{cylinder}$ = 6.3 MPa.

[TABLE 1]

Prior to the density measurements at UVa, the mixtures were validated at BAM by gas chromatography (GC) on a multichannel process analyzer (Siemens MAXUM II, Siemens AG). The GC analysis was conducted following the bracketing procedure proposed in the standard ISO 12963 [69], with two calibration ("bracket") mixtures for each composition that were prepared by the same gravimetric method as for the research mixtures. Further details on the GC validation can be found elsewhere [35]. Here, only propane was analyzed. The results of the GC analysis are reported in Table 2.

[TABLE 2]

## 2.2. Equipment description

The high precision equipment used in this study to obtain the density of the gas mixtures is a single-sinker magnetic suspension coupling densimeter (SSMSD). This technique is founded on the Archimedes' principle which relates the buoyancy force experimented by a sinker of known volume submerged in a fluid with the density of this fluid. It is a primary technique, which means that it works without calibration fluids if an independent determination of the mass and volume of the sinker is available. The first densimeters of this kind were constructed by Kleinrahm et al. [70] in the 1980s, based on the developments of the magnetic suspension coupling of Gast and Robens [71] and Lösch et al. [72], respectively, and equipped with two sinkers. Later in the 1990s, Wagner et al. [73] and Klimeck et al. [74] used a simplified version with a single sinker, which yields higher experimental uncertainty at low densities, but it was also capable of maintaining nearly the same accuracy at moderate and high densities. A detailed schematic diagram of our equipment can be found in another paper [75], together with the current state reported in a recently published article [76]. For an overview of all techniques available for the determination of the density of a fluid, we refer to the comprehensive reports by Wagner et al. and McLinden [77,78].

To sum up briefly, our densimeter has a cylindrical monocrystalline silicon sinker of a mass of (61.59181 ± 0.00012) g and a volume of (26.4440 ± 0.0026) cm$^3$ inside a pressurized cell made of a diamagnetic CuCrZr alloy of specific magnetic susceptibility $\chi_s = -0.025 \cdot 10^{-8}$ m$^3$·kg$^{-1}$ [79]. In a coaxial arrangement with the sinker, there is a sinker support that allows the sinker to be lifted when a permanent magnet placed in the upper edge of the sinker support is attracted by an electronically controlled electromagnet from Rubotherm GmbH. This electromagnet is positioned outside the cell, at ambient pressure, suspended from the hook of an analytical high-precision microbalance (XPE205DR, Mettler Toledo GmbH). With this mounting, the buoyancy force is transmitted without any mechanical contact. The electric power supply to the electromagnet to move the sinker support between the different measuring positions is set according to the feedback signal of a position sensor located on the bottom edge of the sinker support, sufficiently far away from any influence of the magnets. Above the upper pan of the microbalance, an automatically controlled weight changing device from Rubotherm GmbH alternately places two calibrated load compensation masses; one of tantalum, with a mass of (82.0883 ± 0.0002) g and a corresponding volume of (4.9240 ± 0.0005) cm$^3$, and the other of titanium, with a mass of (22.39968 ± 0.00004) g and a volume (4.9706 ± 0.0005) cm$^3$. This setup is within an insulation housing above the measuring cell and rests on a vibration-free structure that also ensures a good leveling of the whole system.

The measuring cell, the sinker, and the gas inlet/outlet ducts are specially made following some precautions to reduce the sorption effect of the gas in the cell materials, which can affect the density determination by up to 0.1 % [80]. For this reason, the use of porous materials, such as elastomeric seals, and the presence of dead volumes and sharp edges in the inner walls was avoided. Another adopted recommendation was to decrease the roughness of the surfaces to the possible minimum by using polished metals.

Two thermostat devices administer the cell to the selected desired set temperature: an outer thermostat that comprises a stainless-steel double-walled cylinder through which the heat-transfer oil from a refrigerating–heating thermostatic bath (JFP50-HE, Julabo GmbH) is circulated, and an inner electrical heating cylinder powered by an electronic controller (MC-E, Julabo GmbH), in direct contact with the measuring cell. Two standard platinum resistance thermometers SPRT-100 are plugged into this controller: the first is placed in the middle of the cell and operates as the temperature-control probe, the second is a safety temperature probe. The temperature inside the cell is recorded as the mean value over a period of time of the readings of an AC resistance bridge (ASL F700, Automatic Systems Laboratory) connected with two SPRT-25 (S1059PJ5X6, Minco Products Inc.) located opposite each other at half the height of the pressurized part of the measuring cell. Another SPRT-25 (162D, Rosemount Inc.) near to the coupling housing is connected to the AC resistance bridge to monitor temperature gradients. The three SPRT-25 were calibrated in our accredited laboratory on the ITS-90 temperature scale [81] with an overall expanded ($k$ = 2) uncertainty of $U(T)$ = 15 mK.

There are two inlet tunnels in the upper and bottom parts of the cell for the inlet and outlet gas ducts. The gas manifold allows for the filling of the gas directly from the supplied bottle and the pressure can be increased after several loadings using a manual piston pump. In addition, a vacuum pump (TRIVAC D8B, Leybold GmbH) with a liquid-nitrogen trap is connected to the manifold to perform measurements in vacuum after finishing each isothermal measurement sequence. The pressure is measured by means of two piezoelectric pressure transducers placed at the top of the measuring cell, covering the ranges (0 to 3) MPa (Digiquartz 2300A-101, Paroscientific Inc.) and (0 to 20) MPa (Digiquartz 43KR-HHT-101, Paroscientific Inc.), respectively. These pressure transducers have also been previously calibrated in our accredited laboratory with an expanded ($k$ = 2) uncertainty of $U_{\mathrm{cal}}(p) = (7.5\cdot10^{-5}(p/\mathrm{MPa}) + 4\cdot10^{-3})$ MPa for the low-pressure transducer and $U_{\mathrm{cal}}(p) = (6.0\cdot10^{-5}(p/\mathrm{MPa}) + 2\cdot10^{-3})$ MPa for the high-pressure transducer.

### 2.3. Density determination

Essentially, the density of the fluid is calculated, according to Archimedes' principle, as the ratio of the buoyant force exerted on the sinker divided by the volume of the sinker. To avoid systematic errors, cancel the influence of the sinker support and auxiliary devices, and decrease the influence of the balance itself in the final result, two compensation masses, made of titanium and tantalum, are used while operating the system in a differential way: zero and measuring positions. The complete procedure is described in detail in [75].

Finally, the density of the fluid $\rho_{\mathrm{fluid}}$ is obtained from the following working equation:

$$\rho_{\mathrm{fluid}} = \frac{\phi_0 m_{\mathrm{s}} + (m_{\mathrm{Ti}} - m_{\mathrm{Ta}}) + \frac{(W_{\mathrm{ZP}} - W_{\mathrm{MP}})}{\alpha}}{V_{\mathrm{s}}(T,p)} \frac{1}{\phi_0} + \frac{\varepsilon_\rho}{\phi_0} \frac{\chi_{\mathrm{s}}}{\chi_{\mathrm{s0}}} \left( \frac{\rho_{\mathrm{s}}}{\rho_0} - \frac{\rho_{\mathrm{fluid}}}{\rho_0} \right) \rho_{\mathrm{fluid}} \quad (1)$$

where the subscripts fluid, s, Ti, Ta, ZP, and MP stand for the fluid, sinker, titanium and tantalum compensation masses, and zero and measuring positions of the magnetic coupling. The terms $m$, $V$, and $\rho$ denote the mass, volume, and density, respectively. $\chi_{\mathrm{s0}} = 10^{-8}$ m$^3$·kg$^{-1}$ and $\rho_0 = 1000$ kg·m$^{-3}$ only work as reducing constants for the specific magnetic susceptibility and density of the fluid, respectively. This equation is obtained by calculating the difference between the balance readings $W$ in two different vertical positions of the magnetic coupling device: in the zero position (ZP), the electromagnet hanging on the lower hook of the balance attracts the permanent magnet without lifting the sinker and the tantalum mass is simultaneously placed on the upper pan of the balance; whereas, in the measuring position (MP), a higher attraction is exerted, now also lifting the sinker, whereas the titanium mass is placed on the pan.

On the one hand, the differential nature of the measuring procedure allows systematic errors in the measurement to be cancelled, as well as the weights of the sinker support, the magnets, and the balance hook; on the other hand, this technique allows the balance to be operated close to its zero, thus avoiding errors associated with the non-linearity of the balance reading. The latter is possible because the difference between the masses of the titanium $m_{\mathrm{Ti}}$ and the tantalum $m_{\mathrm{Ta}}$ compensation masses are selected to be similar to the mass of the silicon sinker. With this fact, their volumes are also nearly the same and the air buoyancy effect on the compensation masses is negligible. In addition, the compensation masses are used to calibrate the balance reading by measuring their masses without lifting the sinker in two previous steps before each measuring point, which enables the calculation of the calibration factor $\alpha$:

$$\alpha = \frac{W_{\mathrm{Ta}} - W_{\mathrm{Ti}}}{m_{\mathrm{Ta}} - m_{\mathrm{Ti}}} \tag{2}$$

where $m_{\mathrm{Ta}}$ and $m_{\mathrm{Ti}}$ are the certified calibrated masses of the tantalum and titanium load compensation masses, respectively.

The distance between the electromagnet and permanent magnet is kept constant by an electronic control unit which triggers the electromagnet with a zero electrical current on average, which avoids heating the surrounding air and the resulting instabilities of the balance reading. However, the position of the permanent magnet inside the measuring cell changes between the zero position and the measuring position, leading to a different effect of the magnetic surrounding elements on the force transmission through the magnetic suspension coupling system. This particular effect is denoted *the force transmission error* and is accounted for by the coupling factor $\phi$. This term accounts for the efficiency of the magnetic suspension coupling: the greater the deviation from unity, the greater the magnitude of the transmission error. Positive values of $(\phi - 1)$ indicate an overall diamagnetic behavior of the surroundings of the magnetic coupling, negative values a paramagnetic one. The coupling factor can be divided into two terms, an apparatus-specific effect $\phi_0$ and a fluid-specific effect $\phi_{\mathrm{fse}}$, thus $\phi = \phi_0 + \phi_{\mathrm{fse}}$[82].

By weighing the sinker in vacuum ($\rho_{\mathrm{fluid}} = 0$), the apparatus-specific effect is determined as:

$$\phi_0 = \frac{-(m_{\mathrm{Ti}} - m_{\mathrm{Ta}}) - \frac{(W_{\mathrm{ZP,vacuum}} - W_{\mathrm{MP,vacuum}})}{\alpha}}{m_{\mathrm{s}}} \tag{3}$$

and the first part of the force transmission error can be corrected. Note that $\phi_0$ depends on the temperature because the strength of the permanent magnet is temperature dependent. Furthermore, small perturbations in the alignment between the electromagnet and the permanent magnet, as well as in the leveling of the balance, cause variations in this value.

The second term, the fluid-specific effect, is mandatory when dealing with paramagnetic fluids, which have a strong temperature-dependent specific magnetic susceptibility $\chi_{\mathrm{s}}$ (note that here, the subscript s stands for *specific* and not for *sinker*). For instance, neglecting this correction on the density of pure oxygen (paramagnetic fluid), measured by a magnetic suspension coupling densimeter, leads to errors of up to 3 % [75]; while the application over measurements of pure nitrogen (diamagnetic fluid) improves the results by less than 0.01 %, thus within the experimental uncertainty. As can been seen from the second right hand term of Equation (1), the fluid-specific effect is also proportional to the apparatus-specific constant $\varepsilon_\rho$, the difference between the densities of the sinker and the fluid, and the density of the fluid itself (a good approximation for evaluating this expression is to consider the density that yields from the Equation (1) with $\phi = \phi_0$, i.e., neglecting the fluid-specific effect). The apparatus-specific constant $\varepsilon_\rho$ of our densimeter has been estimated by two different methods and is reported in our previous work [75]: by measuring the density of a diamagnetic fluid of known magnetic susceptibility, such as nitrogen, with two sinkers of different density at the same temperatures and pressures each time, and by measuring the density with only one sinker, but in a paramagnetic gas, such as oxygen (though standard air can also be used [83]). It was found that the second procedure, considering only a temperature-dependence of the magnetic susceptibility of the paramagnetic pure oxygen, yields more reliable results of $\varepsilon_\rho$ and led to the expression:

$$\varepsilon_\rho(T,\rho) = 8.822 \cdot 10^{-5} + 4.698 \cdot 10^{-8} \cdot \left(\frac{T}{\mathrm{K}} - 293.15\right) - 3.015 \cdot 10^{-8} \cdot \frac{\rho}{(\mathrm{kg{\cdot}m^{-3}})} \quad (4)$$

The specific magnetic susceptibility $\chi_{\mathrm{s,mixture}}$ for the three ($H_2$ + $C_3H_8$) mixtures has been obtained from the mass-based magnetic susceptibility of hydrogen, $\chi_{\mathrm{s,H_2}}$ = –2.49·10$^{-8}$ m$^3$·kg$^{-1}$, and propane, $\chi_{\mathrm{s,C_3H_8}}$ = –1.10·10$^{-8}$ m$^3$·kg$^{-1}$, as:

$$\chi_{\mathrm{s,mixture}}(T) = x_{\mathrm{H_2}} \frac{M_{\mathrm{H_2}}}{M_{\mathrm{mixture}}} \chi_{\mathrm{s,H_2}} + x_{\mathrm{C_3H_8}} \frac{M_{\mathrm{C_3H_8}}}{M_{\mathrm{mixture}}} \chi_{\mathrm{s,C_3H_8}} \quad (5)$$

where $x_{\mathrm{H_2}}$ and $x_{\mathrm{C_3H_8}}$ are the molar fractions and $M_{\mathrm{H_2}}$, $M_{\mathrm{C_3H_8}}$, and $M_{\mathrm{mixture}}$ are the molar masses of the components of the mixture and of the mixture itself, respectively. Note that we have used the specific magnetic susceptibilities in SI units (m$^3$·kg$^{-1}$), but in several cases, the published data found in the literature are in the cgs system of units (cm$^3$·mol$^{-1}$). In these cases, values given in the cgs units must be multiplied by a factor of 4π·10$^{-3}$·[$M^{-1}$·(g·mol$^{-1}$)] to convert them into SI units. The evaluated correction due to the fluid-specific effect in the mixtures measured in this work can be as high as 0.014 kg·m$^{-3}$ at the highest density, $\rho$ = 45.006 kg·m$^{-3}$, being around 0.03 % for all the values. As expected for a mixture with such low magnetic susceptibility, the influence remains well within the experimental uncertainty for every measured coordinate; thus, the medium-specific term of the force transmission error is negligible and could be discarded, in the same manner as done in other works dealing with diamagnetic mixtures [84–86]. In any case, this correction has been considered in this case.

The last unknown variable in Equation (1) is the behavior of the volume of the sinker $V_s(T,p)$ with the temperature and pressure. Under the assumption that a material is isotropic, $V_s(T,p)$ can be evaluated from:

$$V_s(T,p) = V_0(T_0,p_0)\left[1 + 3\alpha(T)(T - T_0) - \frac{3(p-p_0)}{E(T)}(1 - 2\nu(T))\right] \quad (6)$$

where $\alpha$ is the linear thermal expansion coefficient, $E$ is Young's module, $\nu$ is Poisson's ratio, $V_0$ is the sinker volume ~~value~~ specified by the accompanying calibration certificate, and $T_0$ and $p_0$ are the reference state temperature and pressure. The thermal and elastic properties of the silicon of our sinker were obtained from the correlations reported in refs. [87] and [88,89], respectively. Monocrystalline silicon is an isotropic material regarding the thermal expansion, but an anisotropic material concerning the mechanical behavior; thus, the average elastic constants have been used to evaluate the elastic properties $E$ and $\nu$.

### 2.4. Experimental procedure and sorption tests

For six isotherms at $T$ = (250, 275, 300, 325, 350, and 375) K and at pressures between $p$ = (1 to 20) MPa, density data were obtained with the setup described above and executed by starting from a higher pressure down to lower pressures in 1-MPa steps controlled by an automated air-operated valve. The location of the experimental points for the three binary ($H_2$ + $C_3H_8$) mixtures is illustrated in the $p$-$T$ diagrams of Figure 1. The $p$-$T$ range of applicability of the GERG-2008 EoS [50,51] and the main area of interest for the hydrogen economy applications are also represented in Figure 1. The $p$-$T$ area of interest for the hydrogen economy depicted in Figure 1 considers only the processes involved in the hydrogen production, distribution, and storage in depleted gas fields and salt caverns [9,11], but not the high-pressure areas of storage for hydrogen powered vehicles or storage in underground tanks [90,91].

[FIGURE 1]

The calibration factor of the balance, $\alpha$, is determined before and after every single measurement, and the apparatus-specific effect, $\phi_0$, is obtained by a vacuum measurement that is carried out at the end of every isotherm to get the maximum accuracy in the density determination.

This experimental procedure was validated right before and after the measurements of the ($H_2$ + $C_3H_8$) mixtures. The validation is performed by measuring the density of a reference fluid (nitrogen in this case) over the complete operational range of the densimeter. The measured densities of nitrogen were compared to the reference equation of state for nitrogen by Span et al. [92]. The relative deviations of the densities of nitrogen, measured with the densimeter, agreed with those calculated from the reference EoS for nitrogen. In fact, the relative deviations lie within a ± 0.02 % band, and the absolute average deviation (*AAD*) is less than 0.01 %.

As a good practice to minimize possible sorption effects, the evacuated cell was purged several times with fresh sample gas before starting with any isotherm. The apparatus is equipped with a separate inlet-gas in the bottom of the measuring cell and an outlet-gas in the top that can be used for this purpose. Several tests were carried out to discard any distortion in the density results caused by the gas sorption and, hence, a change in the composition of the gas mixtures under study in this work. A continuous recording of the balance mass reading was gathered at the same pressure point for every isotherm investigated for 48 hours, the total duration of recording an isotherm. The magnitude of the relative difference between the first and last record was one order of magnitude lower than the experimental uncertainty at this state point. Several points were also repeated at different times, by ascending and descending pressures, with a repeatability that also remained well within the experimental uncertainty.

## 2.5. Uncertainty of the measurements

The experimental uncertainty of the density measurements, $U(\rho)$, was thoroughly analyzed in a previous study [93]. It was evaluated considering all the terms that contribute to $U(\rho)$, such as the mass readings of the balance and the sinker volume uncertainties (see Section 5.1.3 of [93]) but correcting the experimental density restricted to the apparatus-specific effect. Then, the effect on the uncertainty of the density from the uncertainty of the fluid specific effect was also studied [75], leading to the expression:

$$\frac{U(\rho)}{(\mathrm{kg \cdot m^{-3}})} = \frac{2.5 \cdot 10^{4} \chi_s}{(\mathrm{m^3 \cdot kg^{-1}})} + \frac{1.1 \cdot 10^{-4} \rho}{(\mathrm{kg \cdot m^{-3}})} + 2.3 \cdot 10^{-2} \tag{7}$$

Considering the uncertainty of the pressure and temperature determination of the state point $u(p)$ and $u(T)$ (see Sections 5.1.1 and 5.1.2 of [93]) and the uncertainty of the composition of the mixtures $u(x_i)$ reported in Table 1, the overall expanded ($k$ = 2) uncertainty $U_T(\rho)$ is evaluated from the application of the law of propagation of uncertainty [94] and results in the expression:

$$U_\mathrm{T}(\rho) = 2\left[u(\rho)^2 + \left(\left.\frac{\partial \rho}{\partial p}\right|_{T,x} u(p)\right)^2 + \left(\left.\frac{\partial \rho}{\partial T}\right|_{p,x} u(T)\right)^2 + \sum_i \left(\left.\frac{\partial \rho}{\partial x_i}\right|_{T,p,x_j \neq x_i} u(x_i)\right)^2\right]^{0.5} \tag{8}$$

Partial derivatives were calculated from the GERG-2008 EoS [50,51] using the REFPROP 10.0 software [95,96]. The individual contributions of temperature, pressure, composition, and density to $U_T(\rho)$, together with $U_T(\rho)$, for the three ($H_2$ + $C_3H_8$) mixtures presented in this work are reported in Table 6. The relative expanded ($k$ = 2) uncertainty of the density $U_{T,r}(\rho)$ ranges from 0.11 % for the (0.90 $H_2$ + 0.10 $C_3H_8$) mixture at the highest experimental density up to 1.7 % for the (0.95 $H_2$ + 0.05 $C_3H_8$) at the lowest measured density. The main contribution to $U_{T,r}(\rho)$ originates from the density measurement, between (0.06 to 1.7) %. It is followed by the contributions from pressure and composition which amount to 0.36 % and 0.15 %, respectively, which are significant terms in the range of intermediate to high pressures. The uncertainty of the temperature is a minor term influencing $U_{T,r}(\rho)$ by less than 0.006 %.

[TABLE 3]

## 3. Experimental results

Tables 4, 5, and 6 show the experimental ($p$, $\rho$, $T$) data sets for the three (0.95 $H_2$ + 0.05 $C_3H_8$), (0.90 $H_2$ + 0.10 $C_3H_8$), and (0.83 $H_2$ + 0.17 $C_3H_8$) binary mixtures. The reported temperature, pressure, and density values are the average values calculated from the last ten measurements of a total of thirty. The densities of the experimental points logged in this study span from $\rho$ = 1.316 kg·m$^{-3}$ ($T$ = 375 K, $p$ = 1 MPa, $x_{H_2}$ = 0.95) up to $\rho$ = 45.006 kg·m$^{-3}$ ($T$ = 300 K, $p$ = 20 MPa, $x_{H_2}$ = 0.90). Note that the maximum accessible pressure for the (0.95 $H_2$ + 0.05 $C_3H_8$) mixture at $T$ = 250 K is limited by the dew point at $p$ = 6.9 MPa; for the (0.90 $H_2$ + 0.10 $C_3H_8$) mixture at $T$ = 275 K by the corresponding dew pressure of $p$ = 8.0 MPa; and for the (0.83 $H_2$ + 0.17 $C_3H_8$) mixture at $T$ = 275 K by the dew pressure of $p$ = 3.7 MPa. In addition, pressures of 7 MPa were not exceeded for the system with 0.17 $C_3H_8$ during the preparatory filling steps of the isotherms above the cricondentherm to avoid changes in the mixture composition caused by condensation, because the estimated dew point pressure at an ambient temperature of 296 K is 8.2 MPa.

[TABLES 4 to 6]

Tables 4, 5, and 6 also report the experimental expanded ($k$ = 2) uncertainty of the density measurements $U(\rho)$, stated in absolute units and as a percentage for each data point, and the relative deviations of the experimental densities from the densities given by the AGA8-DC92 EoS [48,49], $\rho_{\text{AGA8-DC92}}$, and the GERG-2008 EoS [50,51], $\rho_{\text{GERG-2008}}$. When using the GERG-2008 EoS, the version that uses the reference equations of state for pure fluids was considered.

In the next Section, the experimental density data are evaluated using the average absolute relative deviation *AARD*, the average relative deviation *Bias*, the root mean square relative deviation *RMS*, and the maximum relative deviation Max*RD*, with respect to the densities calculated from the reference equations of state:

$$AARD = \frac{1}{n}\sum_{i=1}^{n}\left|10^2\,\frac{\rho_{\text{i,exp}}-\rho_{\text{i,EoS}}}{\rho_{\text{i,EoS}}}\right| \tag{9}$$

$$Bias = \frac{1}{n}\sum_{i=1}^{n}\left(10^2\,\frac{\rho_{\text{i,exp}}-\rho_{\text{i,EoS}}}{\rho_{\text{i,EoS}}}\right) \tag{10}$$

$$RMS = \sqrt{\frac{1}{n}\sum_{i=1}^{n}\left(10^2\,\frac{\rho_{\text{i,exp}}-\rho_{\text{i,EoS}}}{\rho_{\text{i,EoS}}}\right)^2} \tag{11}$$

$$\text{Max}RD = \text{maximum}\left(10^2\,\frac{\rho_{\text{i,exp}}-\rho_{\text{i,EoS}}}{\rho_{\text{i,EoS}}}\right) \tag{12}$$

## 4. Discussion

### 4.1. Relative deviations of the experimental data from the reference Helmholtz-type multiparametric equations of state

The relative deviations of experimental density data, $\rho_{\text{exp}}$, with respect to the values calculated from the reference Helmholtz-energy multiparametric mixture models AGA8-DC92 EoS [48,49] and GERG-2008 EoS [50,51], $\rho_{\text{EoS}}$, are calculated using REFPROP 10.0 [95,96] and TREND 5.0 [97] software packages. The version of the GERG-2008 EoS that uses the reference pure-fluid equations for the two constituting compounds $H_2$ and $C_3H_8$, has been used. The results are presented in Tables 3, 4, and 5 and plotted in Figures 2, 3, and 4 for the (0.95 $H_2$ + 0.05 $C_3H_8$), (0.90 $H_2$ + 0.10 $C_3H_8$), and (0.83 $H_2$ + 0.17 $C_3H_8$) mixtures, respectively.

[FIGURES 2 to 4]

The relative deviations tend to increase with the pressure, as well as with the propane content, in all the cases investigated. On the contrary, the higher the temperature, the better the agreement with the two models becomes. The expanded ($k$ = 2) uncertainty of the GERG-2008 model in gas phase density for those systems with only adjusted reducing functions (no departure function) is stated to be about 0.5 % [27,28]. The same value of 0.5 % is recommended for the AGA8-DC92 model [25,26]. These deviations are covered within the joint uncertainties of experimental and EoS-calculated values for the (0.95 $H_2$ + 0.05 $C_3H_8$) mixture for both models. For the (0.90 $H_2$ + 0.10 $C_3H_8$) and (0.83 $H_2$ + 0.17 $C_3H_8$) mixtures, the deviations in density are within the claimed uncertainty of the EoS only when comparing with the AGA8-DC92 model. At pressures higher than 10 MPa and temperatures lower than 325 K, the GERG-2008 EoS fails to reproduce the experimental data for the (0.90 $H_2$ + 0.10 $C_3H_8$) mixture; while, for the (0.83 $H_2$ + 0.17 $C_3H_8$) system, the disagreement is already observed at the highest temperature of 350 K and pressures below 5 MPa. Furthermore, the good agreement below 0.15 % was observed for all isotherms when extrapolating at zero pressure, which is remarkable considering the extremely low density 1.316 kg·m$^{-3}$ for the lowest experimental point, which gives us confidence that no effect that could affect the stability of the mixture composition occurred during the measuring procedure. The minimum estimated uncertainty on average for these systems has been around 0.15 %, so this is our experimental limit. Although this value is only achievable at high pressures of 20 MPa, it is worth pointing out that the extrapolation to zero pressure is also within this threshold.

The overall performance of AGA8-DC92 EoS and GERG-2008 EoS is reported in Table 7 for the ($H_2$ + $C_3H_8$) systems investigated in this work. Clearly, AGA8-DC92 EoS yields systematically better average absolute relative deviations *AARD*, between 0.08 % and 0.28 % as the propane content increases from $x_{\mathrm{C_3H_8}} = (0.05 \text{ to } 0.17)$. These values are two or three times smaller than the corresponding *AARD* from GERG-2008 EoS, with values between 0.21 % and 0.54 %. Moreover, the maximum relative deviations *MaxRD* are also two times lower for the AGA8-DC92 EoS than for the GERG-2008 EoS, ranging from (0.31 to 0.76) % for the former and up to (0.52 to 0.96) % for the latter model. Thus, the use of the AGA8-DC92 mixture model is recommended over the GERG-2008 equation of state when dealing with mixtures containing secondary alkanes with significant amounts of hydrogen, at least until further developments to improve the GERG-2008 have been carried out. That has recently been done with the binary mixtures of (methane, nitrogen, carbon dioxide, and carbon monoxide) + hydrogen [98].

[TABLE 7]

An analysis of the data by Mihara et al. [52], the only other work available in the literature for the density of ($H_2$ + $C_3H_8$) mixtures, shows that the *AARD* values are nearly independent of the propane content of the mixtures studied within the range $x_{\mathrm{C_3H_8}} = (0.16 \text{ to } 0.27)$. The results for the system with $x_{\mathrm{C_3H_8}} = 0.16$ at 298.15 K and maximum pressure of 4 MPa, show an *AARD* = 0.039 % with respect to the AGA8-DC92 EoS and an *AARD* = 0.17 % with respect to the GERG-2008 EoS. These values are two times lower than the corresponding values of this work for the 300 K isotherm and maximum pressure of 5 MPa. Their measurements also show that AGA8-DC92 EoS performs better than GERG-2008 EoS for this binary system, which is in agreement with our findings.

### 4.2. Determination of virial coefficients

The first theoretical model that represents the volumetric behavior of real gases and gives information about the interaction between molecules and came into application is the virial EoS:

$$\frac{p}{RT} = \sum_{k=1}^{N} \frac{B_k}{M^k} \rho_{\mathrm{exp}}^{k} \quad (13)$$

where $R$ is the molar gas constant, and $B_k$ with $k$ = 1, 2, … (and $B_1$ = 1) stands for the second $B_2 = B$, third $B_3 = C$, … virial coefficients, respectively; $B$ originates from molecular pair interactions, $C$ from groups of three molecules, and so on. For a two-component mixture, $B$ and $C$ are expressed in terms of the virial coefficients of the pure components and interaction virial coefficients as:

$$B(T,x) = x_1^2 B_{11}(T) + 2x_1 x_2 B_{12}(T) + x_2^2 B_{22}(T) \tag{14}$$

$$C(T,x) = x_1^3 C_{111}(T) + x_1^2 x_2 C_{112}(T) + x_1 x_2^2 C_{122}(T) + x_2^3 C_{222}(T) \tag{15}$$

where the subscripts 1 and 2 refer to the compounds $H_2$ and $C_3H_8$, respectively.

Here, we applied the non-linear fitting procedure recommended by Cristancho et al. [99]. The first step is to consider the molar mass of the mixture $M$ as another fitted parameter. Then, the data set for every isotherm and composition is fitted to Equation (13) by selecting different truncation orders and maximum experimental density points. The set of truncation order and maximum density that results in an adjusted $M$ value closest to the true gravimetric one will be the best set. Taking this combination, the last step to obtain the final results consists in performing once more the fitting to the virial equation, Equation (13), but now with the true gravimetric value of $M$ instead of the adjusted one. However, for the isotherms at 250 K for both the (0.95 $H_2$ + 0.05 $C_3H_8$) and (0.90 $H_2$ + 0.10 $C_3H_8$) mixtures, the isotherm at 275 K for the (0.90 $H_2$ + 0.10 $C_3H_8$) mixture, as well as for all the measured isotherms of the (0.83 $H_2$ + 0.17 $C_3H_8$) system, it was not possible to find any combination that produced satisfactory values of the adjusted molar mass $M$, so they were discarded from the calculations of the virial coefficients. We assume that this situation is due to the limited maximum pressure experimentally achievable to maintain a homogeneous mixture. For the processed isotherms, a virial expansion truncated up to the third virial coefficient yields the best values when using all the experimental density points for all isotherms except at 350 K for the (0.90 $H_2$ + 0.10 $C_3H_8$) mixture, which was limited up to 22.388 kg·m$^{-3}$ corresponding to 11 MPa.

[TABLE 8]

The final results are reported in Table 8 for the second and third virial coefficients of the mixture, $B$ and $C$, and the second interaction virial coefficients $B_{12}$, calculated from the mixture coefficients, using the coefficients for the pure components from the reference EoS of hydrogen [100] and propane [101]. The values are accompanied by their respective uncertainties, calculated by the Monte Carlo method [102] for the determination of the uncertainty of the coefficients of a fit. The comparison for the second interaction virial coefficient $B_{12}$ with respect to AGA8-DC92 EoS and GERG-2008 EoS is also given. The *RMS* of the residuals amounts to 0.05 and 0.04 % for the $x_{C_3H_8} = 0.05$ and 0.10 mixtures, respectively; a result that is well within $U_{T,r}(\rho)$ and five times lower than the respective values of the deviations with the AGA8-DC92 EoS and GERG-2008 EoS equations.

The experimental mixture second virial coefficients $B$, for the two ($H_2$ + $C_3H_8$) mixtures that could be processed, range from (8.43 to 14.37) cm$^3$·mol$^{-1}$, in contrast with the slightly temperature-dependent and positive $B$ values for pure hydrogen, (14.03 to 15.63) cm$^3$·mol$^{-1}$ and the rather temperature-dependent and negative for pure propane, (−464.4 to −238.4) cm$^3$·mol$^{-1}$. The second virial coefficients account for the intermolecular potential of non-polar non-associating covalent molecules. The effect of the propane is to weaken the long-range attractive instantaneous dipole-induced dipole interactions of hydrogen. As depicted in Figure 5, the interaction second virial coefficients $B_{12}$ from our experiment increase with increasing temperature, as expected, but there is also a slight dependency of the composition apparent. The $B_{12}$ values computed from both AGA8-DC92 EoS and GERG-2008 EoS models are much less composition dependent. Our experimental $B_{12}$ values for both mixtures are fitted to the expression:

$$B_{12} = N_0 + \frac{N_1}{T} \tag{16}$$

with the regressed parameters reported in Table 9 and a root mean square of the residuals *RMS* of below 3.0 cm$^3$·mol$^{-1}$. The $B_{12}$ values from the experiment show a parallel trend similar to the $B_{12}$ estimated from GERG-2008 EoS but located closer to the $B_{12}$ predicted from AGA8-DC92 EoS as illustrated in Figure 6, with a better agreement of $AAD$ = 2.16 $cm^3 \cdot mol^{-1}$ compared to AGA8-DC92 EoS, in contrast to an $AAD$ = 6.73 $cm^3 \cdot mol^{-1}$ with respect to the GERG-2008 EoS. The experimental $B_{12}$ values found in the literature [52,53,103,104] and collected by Dymond et al. [105] show a flatter behavior with the temperature than those from our experimental data and those from both reference equations. Our data compared to the data set by Mihara et al. [52] at temperatures above 300 K show differences below 1.3 $cm^3 \cdot mol^{-1}$. However, our data present increasing discrepancies, as high as 17 $cm^3 \cdot mol^{-1}$ with respect to the Malesinska et al. [104] data point for lower temperatures. Regarding the low-pressure density measurements by Mihara et al. [52], using the values of $B_{12}$ from Equation (16) to estimate the densities yields a root mean square of the relative deviations about 0.1 %, which is within the limit of the $U_{T,r}(\rho)$ = 0.07 % reported by the authors, and better than the predictions of GERG-2008 EoS, but above the differences from AGA8-DC92 EoS.

[FIGURES 5 AND 6, TABLE 9]

### 4.3. Correlation using the PC-SAFT EoS

The only partial success of the adjustment to the virial EoS due to the limiting maximum experimental pressure of the mixture at $x_{C_3H_8} = 0.17$ encouraged us to study the fitting of the whole experimental data sets of this work to the widely used PC-SAFT EoS [55–57]. In addition, and in contrast to the virial EoS, the PC-SAFT model can predict the thermodynamic properties of a fluid mixture in the homogeneous liquid and vapor, and vapor-liquid equilibria regions. The PC-SAFT EoS considers the effect of mixing different fluids due to the change of the free volume, the intermolecular energies, and the molecular orientations on the thermodynamic properties. It is a model explicitly expressed in terms of the Helmholtz free energy $a$ of the fluid mixture as the sum of $a^0$, describing the ideal gas behavior, and $a^{res}$, describing the residual contribution from all the microscopic forces between molecules:

$$a = a^0 + a^{\mathrm{res}} \tag{17}$$

The residual part is conveniently used in a dimensionless reduced form:

$$\tilde{a}^{\mathrm{res}} = \frac{a^{\mathrm{res}}}{N k_{\mathrm{B}} T} \tag{18}$$

where $N$ stands for the total number of molecules, $k_B$ for the Boltzmann constant, and $T$ for the temperature. It consists of three main terms, the hard-chain reference contribution $\tilde{a}^{\mathrm{hc}}$, the dispersion contribution $\tilde{a}^{\mathrm{disp}}$, and the associating contribution $\tilde{a}^{\mathrm{assoc}}$:

$$\tilde{a}^{\mathrm{res}} = \tilde{a}^{\mathrm{hc}} + \tilde{a}^{\mathrm{disp}} + \tilde{a}^{\mathrm{assoc}} \tag{19}$$

$$\tilde{a}^{\mathrm{hc}} = \overline{m}\tilde{a}^{\mathrm{hs}} - \textstyle\sum_i x_i (m_i - 1) \ln g_{ii}^{\mathrm{hs}}(\sigma_{ii}) \tag{20}$$

$$\tilde{a}^{\mathrm{disp}} = -2\pi\rho' I_1(\eta, \overline{m}) \overline{m^2 \varepsilon \sigma^3} - \pi\rho' \overline{m} C_1 I_2(\eta, \overline{m}) \overline{m^2 \varepsilon^2 \sigma^3} \tag{21}$$

where $\tilde{a}^{\mathrm{hc}}$ and $g_{ii}^{\mathrm{hs}}$ are the Helmholtz free energy and radial distribution functions of the hard-sphere fluid, respectively, $\rho'$ stands for the total number density of molecules (number of molecules per unit volume, 1/Å$^3$), and $\eta$ for the packing fraction. $\tilde{a}^{\mathrm{hc}}$ takes into account the hard-spheres contribution in the chain; while $\tilde{a}^{\mathrm{disp}}$ considers the contribution due to the dispersive and repulsive forces between segments. No association contribution was considered as none of our components are associating systems, thus $\tilde{a}^{\mathrm{assoc}}$ was set to zero. The remaining coefficients are different expressions related to the basic molecular parameters for any system $m$, $\sigma$, and $\varepsilon$, which stands for the number, diameter, and energy of or between the segments, respectively. There are also some universal constants of the model that enter the integrals, $I_1$ and $I_2$. $C_1$ is an abbreviation for the compressibility expression. Mean values are denoted by the dash above the magnitude. Further developments were made to account for contributions into the Helmholtz energy from multipole interactions [106,107].

Assuming pairwise additivity, the modeling of a mixture requires the description of interactions between unlike molecules. In the PC-SAFT EoS, this is done using the generalized Lorentz-Berthelot mixing rules to model the interaction size and energy binary parameters:

$$\sigma_{ij} = \frac{1}{2}\left(\sigma_i + \sigma_j\right) \tag{22}$$

$$\varepsilon_{ij} = \left(1 - k_{ij}\right)\sqrt{\varepsilon_i \varepsilon_j} \tag{23}$$

where the subscripts $i$ = 1 and $j$ = 2 refer to $H_2$ and $C_3H_8$, respectively, and $k_{12}$ is a binary interaction parameter. The pure-component adjustable parameters $m$, $\sigma$, and $\varepsilon$ of hydrogen and propane were taken from Senol [108], since they were determined in the same work by fitting proper density data in supercritical conditions, covering the experimental ranges studied in our work. To account for a realistic internuclear distance between the hydrogen atoms, hydrogen was modeled as a non-spherical molecule, with the number of segments per chain less than 1; whereas propane was modeled as a Lennard-Jones chain-like fluid.

The binary interaction parameter $k_{12}$ of ($H_2$ + $C_3H_8$) was determined by a non-linear least square fitting of our experimental densities to the calculated compressibility factors $Z$ from the PC-SAFT EoS, taking the experimental expanded ($k$ = 2) uncertainties as weights:

$$Z = \frac{Mp}{RT\rho} = 1 + Z^{\mathrm{hc}} + Z^{\mathrm{disp}} = 1 + \eta \left.\frac{\partial \tilde{a}^{\mathrm{res}}}{\partial \eta}\right|_{T,x_i} \tag{24}$$

Table 10 lists the pure component parameters used and the results for the temperature-independent binary parameter $k_{12}$ from the regression of the entire density data measured in this work. The residuals for the three ($H_2$ + $C_3H_8$) mixtures with $x_{C_3H_8} = 0.05, 0.10$, and 0.17 are shown in Figure 7. The root mean square *RMS* of the relative residuals are 0.08 % for the (0.95 $H_2$ + 0.05 $C_3H_8$) mixture, 0.09 % for the () mixture, and 0.3 % for the (0.83 $H_2$ + 0.17 $C_3H_8$) mixture; a result that is similar to that obtained from the virial EoS, slightly better than AGA8-DC92 EoS [48,49] deviations and significantly improved compared to the GERG-2008 EoS [27,28] discrepancies. In addition, the comparison with the other data sets of Mihara et al. [52] for these systems shows that the PC-SAFT EoS with the $k_{12}$ value from Table 10 is able to reproduce the measurements better than the virial EoS developed in this work. The results show a good agreement with the AGA8-DC92 EoS evaluations, but clearly outperform GERG-2008 EoS. This is one of the weak points of multiparametric and highly empirical reference Helmholtz equations of state, particularly of GERG-2008 EoS [27,28], because they need a continuous reparameterization with a large databank of wide range and, most importantly, consolidated experimental data sets for many properties to maintain their predictive accuracy, as demonstrated in several works [98,109–111]. On the contrary, more theoretical molecular models also working on Helmholtz energy, such as the PC-SAFT EoS, are much less affected by this issue.

[TABLE 10, FIGURE 7]

If a temperature dependence of $k_{12}$ is assumed, then the regression of the PC-SAFT EoS to the experimental ($p$, $\rho$, $T$) data leads to the expression for the binary interaction parameter of $k_{12}(T/\mathrm{K}) = 34.9/T - 0.0114$, yielding nearly the same overall *RMS* value for the relative residuals of 0.16 %. Therefore, we consider that the temperature-independent $k_{12}$ parameter is sufficient to reproduce our data. Other works that have correlated the SAFT-type EoS to hydrogen-containing systems are available in the literature [61,62].However, as far as we know, this is the first attempt for ($H_2$ + $C_3H_8$) mixtures. On the other hand, if $k_{12}$ = 0, equations (22) and (23) are reduced to the Lorentz-Berthelot combining rules with no fitting to binary experimental data. In this particular case, the comparison of our experimental density data from the evaluations of the PC-SAFT EoS [55–57] yields an overall RMS of the residuals that rises up to 0.29 %, which is above the experimental expanded ($k$ = 2) uncertainty $U_{\mathrm{T,r}}(\rho)$ in most state points, and therefore does not reproduce all the data within its accuracy.

## 5. Conclusions

A high-accuracy single-sinker densimeter, arranged with a magnetic suspension coupling system, was used to obtain density measurements of three binary mixtures of hydrogen and propane, with nominal compositions of (0.95 $H_2$ + 0.05 $C_3H_8$), (0.90 $H_2$ + 0.10 $C_3H_8$), and (0.83 $H_2$ + 0.17 $C_3H_8$), at temperatures between 250 and 375 K and pressures up to 20 MPa. These synthetic mixtures were prepared gravimetrically, with the lowest achievable uncertainty in composition, at the Federal Institute for Materials Research and Testing (BAM) in Berlin, Germany.

Relative deviations between the experimental densities and densities calculated with the AGA8-DC92 EoS are, in most cases, within the ±0.5 % band; which is the estimated uncertainty value of this EoS for this kind of mixtures, except for a limited number of points for the mixture with a higher propane content (17 %) at temperatures of 325 and 350 K, and pressures between 2 and 4 MPa. Relative deviations between the experimental densities and densities calculated with the GERG-2008 EoS are systematically larger. They are within the ±0.5 % band, which is also the estimated uncertainty value of this EoS for this kind of mixtures, for the mixture with 5 % of propane, but deviations are higher than 0.5 % for several isotherms for the mixtures with 10 % and 17 % of propane, especially at low temperatures and high pressures.

The virial equation of state was used to obtain the second $B(T, x)$, the third $C(T, x)$, and the second interaction $B_{12}(T)$ virial coefficients for the mixtures with 5 % and 10 % of propane from the experimental data sets of this work. Virial coefficients for the mixture with 17 % of propane were not obtained, due to the limited number of experimental points measured. The virial equation of state, with these adjusted coefficients, reproduces the recorded density values very well, as can be seen in Table 7.

Finally, this work provides an interaction parameter for hydrogen–propane mixtures optimized for the PC-SAFT EoS [55–57] based on these high-precision density measurements. Obtaining this interaction parameter benefits from the availability of high-pressure measurements, but it is not an essential requirement as it is in the case of virial type equations. The PC-SAFT model [55–57], with the interaction coefficient value estimated here, is able to reproduce our data within their expanded experimental uncertainty range of (0.11 to 1.7) %, as well as the other data sets existing in the literature [52,53] within their standard uncertainty of 0.07 %, better than the multiparameter Helmholtz-type reference equation GERG-2008 [27,28].

The results of these investigations will improve the knowledge of hydrogen-enriched LPG mixtures, for its use in internal combustion engines, and of hydrogen-enriched natural gas mixtures, for the injection of hydrogen in the natural gas grid.

## Acknowledgments

This work was funded by the European Metrology Programme on Innovation and Research (EMPIR, co-funded by the European Union’s Horizon 2020 Research and Innovation Programme and EMPIR Participating States), project 19ENG03/g07 MefHySto, the Ministerio de Economía, Industria y Competitividad, project ENE2017-88474-R, and the Junta de Castilla y León, project VA280P18.

**Tables**

**Table 1.** Purity, supplier, molar mass, and critical parameters of the constituting pure components of the studied mixtures. Molar fractions $x_i$ and expanded ($k$ = 2) uncertainty $U(x_i)$ of the prepared binary ($H_2$ + $C_3H_8$) mixtures. Impurity compounds in the mixtures are marked in italic type.

| Component | Purity / vol-% | Supplier | $M$ / g·mol$^{-1}$ | Critical parameters[a] | |
|---|---|---|---|---|---|
| | | | | $T_c$ / K | $p_c$ / MPa |
| Hydrogen (normal) | 99.9999 | Linde | 2.01588 | 33.145 | 1.2964 |
| Propane | 99.999 | Air Liquide | 44.0956 | 369.89 | 4.2512 |

| Component | (0.95 $H_2$ + 0.05 $C_3H_8$)[b] | | (0.90 $H_2$ + 0.10 $C_3H_8$)[c] | | (0.83 $H_2$ + 0.17 $C_3H_8$)[d] | |
|---|---|---|---|---|---|---|
| | $10^2$ $x_i$ / mol/mol | $10^2$ $U(x_i)$ / mol/mol | $10^2$ $x_i$ / mol/mol | $10^2$ $U(x_i)$ / mol/mol | $10^2$ $x_i$ / mol/mol | $10^2$ $U(x_i)$ / mol/mol |
| Hydrogen (normal) | 95.003 | 0.013 | 90.007 | 0.013 | 83.063 | 0.030 |
| Propane | 4.99650 | 0.00061 | 9.99277 | 0.00059 | 16.9373 | 0.0014 |
| *Nitrogen* | 0.000052 | 0.000055 | 0.000055 | 0.000053 | 0.000034 | 0.000027 |
| *Oxygen* | 0.000025 | 0.000027 | 0.000025 | 0.000026 | 0.000008 | 0.000007 |
| *Carbon dioxide* | 0.000012 | 0.000011 | 0.000014 | 0.000012 | 0.000011 | 0.000010 |
| *Carbon monoxide* | 0.000010 | 0.000011 | 0.000009 | 0.000010 | 0.000002 | 0.000002 |
| *Propylene* | 0.000005 | 0.000006 | 0.000010 | 0.000012 | 0.000017 | 0.000020 |
| Normalized composition without impurities | | | | | | |
| Hydrogen (normal) | 95.003 | 0.013 | 90.007 | 0.013 | 83.063 | 0.030 |
| Propane | 4.99650 | 0.00061 | 9.99278 | 0.00059 | 16.9374 | 0.0014 |

[a] Critical parameters were obtained by using the reference equations of state for hydrogen (normal) [69] and for propane [70] implemented in REFPROP 10.0 software [71].

[b] BAM cylinder no.: 2020-200803.

[c] BAM cylinder no.: 2023-200803.

[d] BAM cylinder no.: 96054 989-210629.

**Table 2.** Results of the gas chromatographic (GC) analysis and relative deviations between gravimetric preparation and GC analysis for the three ($H_2$ + $C_3H_8$) mixtures studied in this work. The results are followed by the gravimetric composition of the employed validation mixtures.

| Component | Concentration | | Relative deviation between gravimetric composition and GC analysis |
|---|---|---|---|
| | $10^2\ x_i$ / mol/mol | $10^2\ U(x_i)$ / mol/mol | % |
| (0.95 $H_2$ + 0.05 $C_3H_8$) BAM no.: 2020-200803 | | | |
| Hydrogen (normal) | n. a.[a] | | |
| Propane | 5.0017 | 0.0033 | 0.105 |
| Validation mixture A BAM no.: 2038-201102 (lower bracket) | | | |
| Hydrogen (normal) | 95.261 | 0.022 | |
| Propane | 4.73927 | 0.00101 | |
| *Nitrogen* | 0.000052 | 0.000055 | |
| *Oxygen* | 0.000025 | 0.000028 | |
| *Carbon dioxide* | 0.000012 | 0.000011 | |
| *Carbon monoxide* | 0.000010 | 0.000011 | |
| *Propylene* | 0.000005 | 0.000005 | |
| Validation mixture B BAM no.: 2039-201102 (upper bracket) | | | |
| Hydrogen (normal) | 94.744 | 0.022 | |
| Propane | 5.25608 | 0.00102 | |
| *Nitrogen* | 0.000053 | 0.000055 | |
| *Oxygen* | 0.000025 | 0.000028 | |
| *Carbon dioxide* | 0.000012 | 0.000011 | |
| *Carbon monoxide* | 0.000009 | 0.000011 | |
| *Propylene* | 0.000005 | 0.000006 | |
| (0.90 $H_2$ + 0.10 $C_3H_8$) BAM no.: 2023-200803 | | | |
| Hydrogen (normal) | n. a.[a] | | |
| Propane | 9.9952 | 0.0060 | 0.024 |
| Validation mixture C BAM no.: 8078-201026 (lower bracket) | | | |
| Hydrogen (normal) | 90.524 | 0.018 | |
| Propane | 9.47592 | 0.00081 | |
| *Nitrogen* | 0.000055 | 0.000053 | |
| *Oxygen* | 0.000025 | 0.000026 | |
| *Carbon dioxide* | 0.000014 | 0.000012 | |
| *Carbon monoxide* | 0.000009 | 0.000010 | |
| *Propylene* | 0.000009 | 0.000011 | |
| Validation mixture D BAM no.: 2045-201026 (upper bracket) | | | |
| Hydrogen (normal) | 89.504 | 0.018 | |
| Propane | 10.49563 | 0.00081 | |
| *Nitrogen* | 0.000055 | 0.000053 | |
| *Oxygen* | 0.000025 | 0.000026 | |
| *Carbon dioxide* | 0.000014 | 0.000012 | |
| *Carbon monoxide* | 0.000009 | 0.000010 | |
| *Propylene* | 0.000010 | 0.000012 | |
| ($0.83\ H_2 + 0.17\ C_3H_8$) BAM no.: 96054 989-210629 | | | |
| Hydrogen (normal) | n. a.[a] | | |
| Propane | 16.9467 | 0.0231 | 0.055 |
| Validation mixture E BAM no.: 96054988-210629 (lower bracket) | | | |
| Hydrogen (normal) | 83.883 | 0.030 | |
| Propane | 16.11672 | 0.00136 | |
| *Nitrogen* | 0.000033 | 0.000027 | |
| *Oxygen* | 0.000008 | 0.000007 | |
| *Carbon dioxide* | 0.000010 | 0.000010 | |
| *Carbon monoxide* | 0.000002 | 0.000002 | |
| *Propylene* | 0.000016 | 0.000019 | |

Validation mixture F BAM no.: 96054990-210629 (upper bracket)

| | | |
|---|---|---|
| Hydrogen (normal) | 82.186 | 0.030 |
| Propane | 17.81393 | 0.00139 |
| *Nitrogen* | 0.000034 | 0.000028 |
| *Oxygen* | 0.000009 | 0.000007 |
| *Carbon dioxide* | 0.000011 | 0.000011 |
| *Carbon monoxide* | 0.000002 | 0.000002 |
| *Propylene* | 0.000018 | 0.000021 |

[a] not analyzed (balance gas).

**Table 3.** Contributions to the overall expanded ($k$ = 2) relative uncertainty in density, $U_{T,r}(\rho)$, for the three ($H_2$ + $C_3H_8$) mixtures studied in this work.

| Source | Estimation ($k$ = 2) | Units | Contribution to the uncertainty of density ($k$ = 2) | |
|---|---|---|---|---|
| | | | $kg \cdot m^{-3}$ | % |
| | | ($0.95\ H_2 + 0.05\ C_3H_8$) | | |
| Temperature, $T$ | 0.015 | K | < 0.0017 | < 0.0061 |
| Pressure, $p$ | < 0.005 | MPa | (0.00031–0.0073) | (0.012–0.094) |
| Composition, $x_i$ | < 0.0004 | $mol \cdot mol^{-1}$ | < 0.046 | < 0.14 |
| Density, $\rho$ | (0.023–0.026) | $kg \cdot m^{-3}$ | (0.023–0.026) | (0.082–1.7) |
| Sum | | | (0.023–0.053) | (0.17–1.7) |
| | | ($0.90\ H_2 + 0.10\ C_3H_8$) | | |
| Temperature, $T$ | 0.015 | K | < 0.0023 | < 0.0058 |
| Pressure, $p$ | < 0.005 | MPa | (0.00047–0.010) | (0.012–0.10) |
| Composition, $x_i$ | < 0.0004 | $mol \cdot mol^{-1}$ | < 0.041 | < 0.092 |
| Density, $\rho$ | (0.023–0.028) | $kg \cdot m^{-3}$ | (0.023–0.028) | (0.062–1.2) |
| Sum | | | (0.023–0.051) | (0.11–1.2) |
| | | ($0.83\ H_2 + 0.17\ C_3H_8$) | | |
| Temperature, $T$ | 0.015 | K | < 0.0016 | < 0.0058 |
| Pressure, $p$ | < 0.005 | MPa | (0.00070–0.015) | (0.012–0.36) |
| Composition, $x_i$ | < 0.0004 | $mol \cdot mol^{-1}$ | < 0.048 | < 0.15 |
| Density, $\rho$ | (0.023–0.027) | $kg \cdot m^{-3}$ | (0.023–0.027) | (0.083–0.79) |
| Sum | | | (0.024–0.057) | (0.18–0.80) |

**Table 4.** Experimental ($p$, $\rho_{exp}$, $T$) measurements for the binary (0.95 $H_2$ + 0.05 $C_3H_8$) mixture, absolute and relative expanded ($k$ = 2) uncertainty in density, $U(\rho_{exp})$, and relative deviations from the density given by the AGA8-DC92 EoS [48,49], $\rho_{AGA8\text{-}DC92}$, and the GERG-2008 EoS [50,51], $\rho_{GERG\text{-}2008}$.

| $T$ / K[(a)] | $p$ / MPa[(b)] | $\rho_{exp}$ / kg·m$^{-3}$[(c)] | $U(\rho_{exp})$ / kg·m$^{-3}$ | $10^2$ $U(\rho_{exp})/\rho_{exp}$ | $10^2$ ($\rho_{exp}$ − $\rho_{AGA8\text{-}DC92}$)/$\rho_{AGA8\text{-}DC92}$ | $10^2$ ($\rho_{exp}$ − $\rho_{GERG\text{-}2008}$)/$\rho_{GERG\text{-}2008}$ |
|---|---|---|---|---|---|---|
| 250.00 K | | | | | | |
| 250.144 | 3.9950 | 7.746 | 0.024 | 0.305 | −0.07 | −0.33 |
| 250.138 | 3.00119 | 5.856 | 0.023 | 0.400 | 0.04 | −0.17 |
| 250.140 | 1.99826 | 3.922 | 0.023 | 0.592 | 0.10 | −0.03 |
| 250.141 | 1.00086 | 1.973 | 0.023 | 1.165 | 0.05 | −0.02 |
| 275.00 K | | | | | | |
| 275.101 | 19.9863 | 32.105 | 0.026 | 0.082 | −0.31 | −0.52 |
| 275.105 | 19.0222 | 30.737 | 0.026 | 0.085 | n. a.[d] | −0.52 |
| 275.105 | 18.0194 | 29.298 | 0.026 | 0.089 | n. a.[d] | −0.52 |
| 275.106 | 17.0222 | 27.843 | 0.026 | 0.093 | n. a.[d] | −0.52 |
| 275.104 | 16.0576 | 26.423 | 0.026 | 0.098 | n. a.[d] | −0.50 |
| 275.107 | 15.0115 | 24.856 | 0.026 | 0.103 | n. a.[d] | −0.50 |
| 275.108 | 13.9927 | 23.313 | 0.025 | 0.109 | n. a.[d] | −0.48 |
| 275.107 | 13.0019 | 21.788 | 0.025 | 0.116 | n. a.[d] | −0.48 |
| 275.108 | 11.9886 | 20.212 | 0.025 | 0.124 | n. a.[d] | −0.46 |
| 275.109 | 11.0148 | 18.677 | 0.025 | 0.133 | n. a.[d] | −0.43 |
| 275.105 | 9.9937 | 17.048 | 0.025 | 0.145 | n. a.[d] | −0.40 |
| 275.107 | 9.0033 | 15.445 | 0.025 | 0.159 | n. a.[d] | −0.38 |
| 275.103 | 7.9996 | 13.802 | 0.024 | 0.176 | −0.05 | −0.35 |
| 275.105 | 7.0006 | 12.148 | 0.024 | 0.199 | −0.02 | −0.30 |
| 275.110 | 5.8629 | 10.237 | 0.024 | 0.234 | −0.02 | −0.26 |
| 275.106 | 4.9994 | 8.776 | 0.024 | 0.271 | 0.04 | −0.17 |
| 275.111 | 3.9987 | 7.054 | 0.024 | 0.334 | < 0.01 | −0.18 |
| 275.121 | 3.00167 | 5.324 | 0.023 | 0.439 | 0.03 | −0.10 |
| 275.122 | 2.00569 | 3.577 | 0.023 | 0.648 | 0.08 | −0.02 |
| 275.127 | 1.00601 | 1.805 | 0.023 | 1.272 | 0.17 | 0.12 |
| 300.00 K | | | | | | |
| 300.040 | 18.8654 | 28.101 | 0.026 | 0.092 | −0.20 | −0.35 |
| 300.039 | 17.9825 | 26.922 | 0.026 | 0.096 | −0.18 | −0.35 |
| 300.040 | 17.0116 | 25.611 | 0.026 | 0.100 | −0.16 | −0.35 |
| 300.041 | 15.9918 | 24.215 | 0.026 | 0.105 | −0.15 | −0.34 |
| 300.041 | 15.0138 | 22.861 | 0.025 | 0.111 | −0.13 | −0.34 |
| 300.041 | 14.0148 | 21.461 | 0.025 | 0.117 | −0.11 | −0.33 |
| 300.040 | 12.9964 | 20.016 | 0.025 | 0.125 | −0.10 | −0.32 |
| 300.038 | 11.9824 | 18.559 | 0.025 | 0.134 | −0.08 | −0.31 |
| 300.043 | 10.9846 | 17.109 | 0.025 | 0.144 | −0.07 | −0.30 |
| 300.041 | 9.9994 | 15.660 | 0.025 | 0.157 | −0.05 | −0.28 |
| 300.041 | 9.0018 | 14.175 | 0.024 | 0.172 | −0.04 | −0.26 |
| 300.040 | 8.0037 | 12.673 | 0.024 | 0.191 | −0.02 | −0.23 |
| 300.058 | 7.0026 | 11.149 | 0.024 | 0.216 | −0.01 | −0.20 |
| 300.059 | 6.0028 | 9.609 | 0.024 | 0.248 | 0.01 | −0.16 |
| 300.043 | 5.0001 | 8.048 | 0.024 | 0.294 | 0.03 | −0.12 |
| 300.044 | 3.9969 | 6.469 | 0.024 | 0.363 | 0.06 | −0.07 |
| 300.067 | 3.00305 | 4.888 | 0.023 | 0.477 | 0.12 | 0.03 |
| 300.066 | 2.02112 | 3.308 | 0.023 | 0.700 | 0.17 | 0.10 |
| 300.065 | 1.01083 | 1.662 | 0.023 | 1.381 | 0.10 | 0.07 |
| | | | 325.00 K | | | |
| 325.057 | 19.7442 | 27.114 | 0.026 | 0.095 | −0.21 | −0.30 |
| 325.057 | 18.9683 | 26.158 | 0.026 | 0.098 | −0.20 | −0.30 |
| 325.055 | 17.9723 | 24.919 | 0.026 | 0.103 | −0.18 | −0.30 |
| 325.054 | 17.0041 | 23.699 | 0.025 | 0.107 | −0.17 | −0.30 |
| 325.069 | 16.0058 | 22.428 | 0.025 | 0.113 | −0.15 | −0.29 |
| 325.069 | 14.9986 | 21.131 | 0.025 | 0.119 | −0.14 | −0.29 |
| 325.064 | 13.9909 | 19.818 | 0.025 | 0.126 | −0.13 | −0.29 |
| 325.070 | 12.9900 | 18.499 | 0.025 | 0.134 | −0.12 | −0.28 |
| 325.071 | 11.9965 | 17.176 | 0.025 | 0.144 | −0.10 | −0.27 |
| 325.070 | 10.9876 | 15.817 | 0.025 | 0.155 | −0.09 | −0.25 |
| 325.070 | 10.0053 | 14.478 | 0.024 | 0.169 | −0.08 | −0.24 |
| 325.072 | 8.9935 | 13.083 | 0.024 | 0.185 | −0.07 | −0.23 |
| 325.072 | 7.9968 | 11.695 | 0.024 | 0.206 | −0.05 | −0.20 |
| 325.071 | 6.9953 | 10.285 | 0.024 | 0.233 | −0.04 | −0.18 |
| 325.072 | 5.9960 | 8.862 | 0.024 | 0.268 | −0.03 | −0.16 |
| 325.070 | 5.0006 | 7.430 | 0.024 | 0.318 | −0.01 | −0.12 |
| 325.069 | 3.9980 | 5.971 | 0.023 | 0.393 | −0.01 | −0.10 |
| 325.070 | 2.98748 | 4.488 | 0.023 | 0.519 | 0.06 | −0.01 |
| 325.068 | 1.99896 | 3.019 | 0.023 | 0.765 | 0.11 | 0.06 |
| 325.068 | 0.99884 | 1.519 | 0.023 | 1.510 | 0.29 | 0.26 |
| | | | 350.00 K | | | |
| 350.066 | 19.2650 | 24.735 | 0.026 | 0.103 | −0.11 | −0.20 |
| 350.067 | 17.9829 | 23.240 | 0.025 | 0.109 | −0.09 | −0.21 |
| 350.066 | 16.9979 | 22.078 | 0.025 | 0.115 | −0.09 | −0.21 |
| 350.063 | 15.9490 | 20.828 | 0.025 | 0.121 | −0.08 | −0.21 |
| 350.064 | 15.0067 | 19.692 | 0.025 | 0.127 | −0.07 | −0.21 |
| 350.065 | 14.0060 | 18.473 | 0.025 | 0.135 | −0.06 | −0.20 |
| 350.066 | 13.0092 | 17.246 | 0.025 | 0.143 | −0.05 | −0.20 |
| 350.064 | 11.9817 | 15.967 | 0.025 | 0.154 | −0.04 | −0.19 |
| 350.065 | 10.9938 | 14.725 | 0.024 | 0.166 | −0.03 | −0.18 |
| 350.066 | 9.9981 | 13.459 | 0.024 | 0.181 | −0.03 | −0.17 |
| 350.065 | 8.9952 | 12.172 | 0.024 | 0.198 | −0.02 | −0.15 |
| 350.066 | 7.9913 | 10.869 | 0.024 | 0.221 | −0.01 | −0.13 |
| 350.067 | 6.9928 | 9.559 | 0.024 | 0.250 | < 0.01 | −0.12 |
| 350.067 | 5.9988 | 8.241 | 0.024 | 0.288 | 0.01 | −0.10 |
| 350.067 | 4.9965 | 6.899 | 0.024 | 0.341 | 0.01 | −0.08 |
| 350.067 | 3.9963 | 5.546 | 0.023 | 0.422 | 0.02 | −0.05 |
| 350.065 | 2.98028 | 4.157 | 0.023 | 0.559 | 0.02 | −0.03 |
| 350.075 | 2.00505 | 2.810 | 0.023 | 0.821 | 0.04 | < 0.01 |
| 350.079 | 1.01417 | 1.429 | 0.023 | 1.604 | 0.08 | 0.06 |
| | | | 375.00 K | | | |
| 375.061 | 19.8833 | 23.858 | 0.025 | 0.107 | 0.03 | −0.11 |
| 375.061 | 18.9806 | 22.875 | 0.025 | 0.111 | 0.03 | −0.11 |
| 375.060 | 18.0035 | 21.801 | 0.025 | 0.116 | 0.04 | −0.11 |
| 375.060 | 16.9679 | 20.650 | 0.025 | 0.122 | 0.04 | −0.11 |
| 375.059 | 15.9815 | 19.544 | 0.025 | 0.128 | 0.04 | −0.11 |
| 375.060 | 14.9827 | 18.411 | 0.025 | 0.135 | 0.04 | −0.11 |
| 375.059 | 13.9791 | 17.263 | 0.025 | 0.143 | 0.05 | −0.11 |
| 375.055 | 12.9907 | 16.120 | 0.025 | 0.153 | 0.05 | −0.10 |
| 375.061 | 11.9848 | 14.945 | 0.024 | 0.164 | 0.06 | −0.10 |
| 375.062 | 10.9973 | 13.781 | 0.024 | 0.177 | 0.07 | −0.08 |
| 375.062 | 9.9991 | 12.591 | 0.024 | 0.192 | 0.07 | −0.07 |
| 375.060 | 9.0018 | 11.391 | 0.024 | 0.211 | 0.08 | −0.06 |
| 375.060 | 7.9970 | 10.169 | 0.024 | 0.235 | 0.08 | −0.05 |
| 375.061 | 6.9935 | 8.937 | 0.024 | 0.266 | 0.08 | −0.03 |
| 375.060 | 5.9986 | 7.702 | 0.024 | 0.307 | 0.08 | −0.02 |
| 375.059 | 4.9994 | 6.451 | 0.023 | 0.364 | 0.08 | −0.01 |
| 375.060 | 3.9998 | 5.187 | 0.023 | 0.450 | 0.09 | 0.02 |
| 375.062 | 2.98885 | 3.894 | 0.023 | 0.596 | 0.07 | 0.02 |
| 375.059 | 1.99908 | 2.617 | 0.023 | 0.881 | 0.10 | 0.06 |
| 375.060 | 0.99920 | 1.316 | 0.023 | 1.741 | 0.22 | 0.20 |

(a) $U(T)$ = 15 mK

(b) $\frac{U(p>3)}{\mathrm{MPa}} = 75 \cdot 10^{-6} \cdot \frac{p}{\mathrm{MPa}} + 3.5 \cdot 10^{-3}$; $\frac{U(p<3)}{\mathrm{MPa}} = 60 \cdot 10^{-6} \cdot \frac{p}{\mathrm{MPa}} + 1.7 \cdot 10^{-4}$

(c) $\frac{U(\rho)}{\mathrm{kg \cdot m^{-3}}} = 2.5 \cdot 10^{4} \frac{\chi_{\mathrm{s}}}{\mathrm{m^3 kg^{-1}}} + 1.1 \cdot 10^{-4} \cdot \frac{\rho}{\mathrm{kg \cdot m^{-3}}} + 2.3 \cdot 10^{-2}$

(d) not available, because REFPROP 10.0 does not calculate a converging value of the density at this state point using the AGA8-DC92 EoS [48,49].

**Table 5.** Experimental ($p$, $\rho_{exp}$, $T$) measurements for the binary (0.90 $H_2$ + 0.10 $C_3H_8$) mixture, absolute and relative expanded ($k$ = 2) uncertainty in density, $U(\rho_{exp})$, and relative deviations from the density given by the AGA8-DC92 EoS [48,49], $\rho_{AGA8\text{-}DC92}$, and the GERG-2008 EoS [50,51], $\rho_{GERG\text{-}2008}$.

| $T$ / K[(a)] | $p$ / MPa[(b)] | $\rho_{exp}$ / kg·m$^{-3}$[(c)] | $U(\rho_{exp})$ / kg·m$^{-3}$ | $10^2\ U(\rho_{exp})/\rho_{exp}$ | $10^2\ (\rho_{exp} - \rho_{AGA8\text{-}DC92})/\rho_{AGA8\text{-}DC92}$ | $10^2\ (\rho_{exp} - \rho_{GERG\text{-}2008})/\rho_{GERG\text{-}2008}$ |
|---|---|---|---|---|---|---|
| 275.00 K | | | | | | |
| 275.111 | 6.0892 | 16.147 | 0.025 | 0.153 | n. a.[d] | −0.90 |
| 275.111 | 4.9931 | 13.306 | 0.024 | 0.183 | n. a.[d] | −0.76 |
| 275.117 | 3.9949 | 10.694 | 0.024 | 0.225 | n. a.[d] | −0.61 |
| 275.115 | 2.98456 | 8.024 | 0.024 | 0.296 | −0.22 | −0.46 |
| 275.115 | 1.99549 | 5.388 | 0.023 | 0.435 | −0.12 | −0.28 |
| 275.114 | 0.99615 | 2.700 | 0.023 | 0.857 | −0.05 | −0.13 |
| 300.00 K | | | | | | |
| 300.067 | 19.8477 | 45.006 | 0.028 | 0.062 | −0.47 | −0.91 |
| 300.066 | 19.0154 | 43.325 | 0.028 | 0.064 | −0.42 | −0.90 |
| 300.069 | 17.8303 | 40.898 | 0.027 | 0.067 | −0.34 | −0.88 |
| 300.068 | 16.8732 | 38.909 | 0.027 | 0.070 | −0.25 | −0.87 |
| 300.069 | 15.9720 | 37.014 | 0.027 | 0.073 | −0.23 | −0.85 |
| 300.070 | 14.9410 | 34.818 | 0.027 | 0.077 | −0.21 | −0.83 |
| 300.071 | 13.9938 | 32.775 | 0.027 | 0.081 | −0.19 | −0.81 |
| 300.070 | 13.0096 | 30.627 | 0.026 | 0.086 | −0.17 | −0.78 |
| 300.070 | 11.9950 | 28.386 | 0.026 | 0.092 | −0.16 | −0.74 |
| 300.069 | 10.9831 | 26.125 | 0.026 | 0.099 | −0.14 | −0.70 |
| 300.069 | 9.9922 | 23.884 | 0.026 | 0.107 | −0.13 | −0.66 |
| 300.070 | 8.9990 | 21.613 | 0.025 | 0.117 | −0.11 | −0.61 |
| 300.072 | 7.9917 | 19.284 | 0.025 | 0.130 | −0.10 | −0.56 |
| 300.073 | 6.9932 | 16.953 | 0.025 | 0.146 | −0.08 | −0.50 |
| 300.074 | 5.9993 | 14.608 | 0.024 | 0.168 | −0.06 | −0.43 |
| 300.074 | 4.9947 | 12.216 | 0.024 | 0.198 | −0.04 | −0.36 |
| 300.075 | 3.9956 | 9.814 | 0.024 | 0.244 | −0.02 | −0.27 |
| 300.072 | 2.99383 | 7.388 | 0.024 | 0.320 | 0.05 | −0.14 |
| 300.074 | 1.99655 | 4.947 | 0.023 | 0.473 | 0.07 | −0.05 |
| 300.072 | 0.99705 | 2.480 | 0.023 | 0.932 | 0.11 | 0.05 |
| 325.00 K | | | | | | |
| 325.078 | 19.8389 | 41.479 | 0.028 | 0.066 | −0.43 | −0.90 |
| 325.079 | 18.8453 | 39.614 | 0.027 | 0.069 | −0.41 | −0.89 |
| 325.079 | 17.9850 | 37.980 | 0.027 | 0.071 | −0.39 | −0.88 |
| 325.074 | 17.0270 | 36.141 | 0.027 | 0.075 | −0.38 | −0.87 |
| 325.071 | 15.9754 | 34.097 | 0.027 | 0.078 | −0.36 | −0.86 |
| 325.073 | 15.0159 | 32.209 | 0.026 | 0.082 | −0.34 | −0.84 |
| 325.078 | 13.9986 | 30.185 | 0.026 | 0.087 | −0.33 | −0.82 |
| 325.060 | 12.9783 | 28.129 | 0.026 | 0.093 | −0.32 | −0.80 |
| 325.073 | 12.0062 | 26.150 | 0.026 | 0.099 | −0.30 | −0.77 |
| 325.076 | 10.9944 | 24.067 | 0.026 | 0.106 | −0.28 | −0.73 |
| 325.079 | 9.9974 | 21.990 | 0.025 | 0.115 | −0.27 | −0.70 |
| 325.081 | 9.0054 | 19.902 | 0.025 | 0.126 | −0.26 | −0.66 |
| 325.084 | 7.9975 | 17.757 | 0.025 | 0.140 | −0.25 | −0.62 |
| 325.085 | 6.9975 | 15.610 | 0.025 | 0.158 | −0.23 | −0.57 |
| 325.087 | 5.9947 | 13.433 | 0.024 | 0.181 | −0.22 | −0.52 |
| 325.080 | 4.9964 | 11.246 | 0.024 | 0.214 | −0.22 | −0.47 |
| 325.082 | 3.9979 | 9.038 | 0.024 | 0.264 | −0.21 | −0.41 |
| 325.087 | 2.98325 | 6.776 | 0.024 | 0.348 | −0.16 | −0.31 |
| 325.084 | 1.99706 | 4.554 | 0.023 | 0.513 | −0.15 | −0.25 |
| 325.087 | 0.99678 | 2.283 | 0.023 | 1.011 | −0.08 | −0.13 |
| | | | 350.00 K | | | |
| 350.078 | 19.8437 | 38.648 | 0.027 | 0.070 | −0.17 | −0.57 |
| 350.078 | 18.9198 | 37.023 | 0.027 | 0.073 | −0.16 | −0.57 |
| 350.079 | 17.9813 | 35.355 | 0.027 | 0.076 | −0.14 | −0.57 |
| 350.077 | 16.9522 | 33.505 | 0.027 | 0.080 | −0.13 | −0.56 |
| 350.075 | 15.9686 | 31.718 | 0.026 | 0.083 | −0.12 | −0.55 |
| 350.073 | 15.0105 | 29.958 | 0.026 | 0.088 | −0.11 | −0.53 |
| 350.077 | 14.0070 | 28.095 | 0.026 | 0.093 | −0.10 | −0.52 |
| 350.076 | 12.9927 | 26.191 | 0.026 | 0.099 | −0.09 | −0.50 |
| 350.077 | 11.9851 | 24.278 | 0.026 | 0.105 | −0.08 | −0.47 |
| 350.077 | 10.9985 | 22.388 | 0.025 | 0.113 | −0.06 | −0.44 |
| 350.077 | 9.9919 | 20.436 | 0.025 | 0.123 | −0.05 | −0.41 |
| 350.078 | 8.9948 | 18.484 | 0.025 | 0.135 | −0.04 | −0.38 |
| 350.078 | 7.9929 | 16.501 | 0.025 | 0.150 | −0.04 | −0.35 |
| 350.075 | 6.9938 | 14.505 | 0.024 | 0.169 | −0.03 | −0.31 |
| 350.078 | 6.0006 | 12.501 | 0.024 | 0.194 | −0.03 | −0.28 |
| 350.079 | 4.9978 | 10.457 | 0.024 | 0.230 | −0.03 | −0.24 |
| 350.077 | 3.9965 | 8.398 | 0.024 | 0.283 | −0.04 | −0.21 |
| 350.079 | 2.98393 | 6.298 | 0.024 | 0.374 | −0.02 | −0.15 |
| 350.076 | 1.99713 | 4.233 | 0.023 | 0.551 | −0.02 | −0.10 |
| 350.077 | 0.99765 | 2.123 | 0.023 | 1.086 | −0.03 | −0.07 |
| | | | 375.00 K | | | |
| 375.070 | 19.9362 | 36.277 | 0.027 | 0.074 | −0.10 | −0.49 |
| 375.068 | 18.9834 | 34.705 | 0.027 | 0.077 | −0.09 | −0.49 |
| 375.066 | 18.0155 | 33.090 | 0.027 | 0.080 | −0.09 | −0.49 |
| 375.065 | 17.0080 | 31.393 | 0.026 | 0.084 | −0.09 | −0.49 |
| 375.069 | 16.0005 | 29.677 | 0.026 | 0.088 | −0.08 | −0.48 |
| 375.072 | 14.9864 | 27.932 | 0.026 | 0.093 | −0.07 | −0.47 |
| 375.072 | 13.9957 | 26.209 | 0.026 | 0.098 | −0.07 | −0.46 |
| 375.072 | 12.9740 | 24.415 | 0.026 | 0.105 | −0.06 | −0.44 |
| 375.072 | 11.9955 | 22.678 | 0.025 | 0.112 | −0.06 | −0.42 |
| 375.069 | 10.9934 | 20.882 | 0.025 | 0.121 | −0.05 | −0.40 |
| 375.068 | 9.9893 | 19.064 | 0.025 | 0.131 | −0.05 | −0.37 |
| 375.067 | 8.9827 | 17.222 | 0.025 | 0.144 | −0.05 | −0.35 |
| 375.066 | 7.9987 | 15.404 | 0.025 | 0.160 | −0.05 | −0.33 |
| 375.067 | 6.9881 | 13.519 | 0.024 | 0.180 | −0.05 | −0.30 |
| 375.066 | 5.9947 | 11.648 | 0.024 | 0.207 | −0.05 | −0.28 |
| 375.066 | 4.9911 | 9.741 | 0.024 | 0.246 | −0.06 | −0.25 |
| 375.068 | 3.9951 | 7.830 | 0.024 | 0.303 | −0.08 | −0.23 |
| 375.065 | 2.98366 | 5.875 | 0.023 | 0.400 | −0.05 | −0.16 |
| 375.068 | 1.99520 | 3.946 | 0.023 | 0.590 | −0.03 | −0.11 |
| 375.065 | 0.99552 | 1.977 | 0.023 | 1.166 | −0.02 | −0.06 |

(a) $U(T)$ = 15 mK

(b) $\frac{U(p>3)}{\text{MPa}} = 75 \cdot 10^{-6} \cdot \frac{p}{\text{MPa}} + 3.5 \cdot 10^{-3}; \frac{U(p<3)}{\text{MPa}} = 60 \cdot 10^{-6} \cdot \frac{p}{\text{MPa}} + 1.7 \cdot 10^{-4}$

(c) $\frac{U(\rho)}{\text{kg·m}^{-3}} = 2.5 \cdot 10^{4} \frac{\chi_\text{s}}{\text{m}^3\text{kg}^{-1}} + 1.1 \cdot 10^{-4} \cdot \frac{\rho}{\text{kg·m}^{-3}} + 2.3 \cdot 10^{-2}$

[d] not available, because REFPROP 10.0 does not calculate a converging value of the density at this state point using the AGA8-DC92 EoS [48,49].

**Table 6.** Experimental ($p$, $\rho_{exp}$, $T$) measurements for the binary (0.83 $H_2$ + 0.17 $C_3H_8$) mixture, absolute and relative expanded ($k$ = 2) uncertainty in density, $U(\rho_{exp})$, and relative deviations from the density given by the AGA8-DC92 EoS [48,49], $\rho_{AGA8\text{-}DC92}$, and the GERG-2008 EoS [50,51], $\rho_{GERG\text{-}2008}$.

| $T$ / K[(a)] | $p$ / MPa[(b)] | $\rho_{exp}$ / kg·m$^{-3}$[(c)] | $U(\rho_{exp})$ / kg·m$^{-3}$ | $10^2$ $U(\rho_{exp})/\rho_{exp}$ | $10^2$ ($\rho_{exp}$ − $\rho_{AGA8\text{-}DC92}$)/$\rho_{AGA8\text{-}DC92}$ | $10^2$ ($\rho_{exp}$ − $\rho_{GERG\text{-}2008}$)/$\rho_{GERG\text{-}2008}$ |
|---|---|---|---|---|---|---|
| | | | 275.00 K | | | |
| 275.130 | 3.00990 | 12.060 | 0.024 | 0.201 | n. a.[d] | −0.40 |
| 275.136 | 2.99302 | 11.988 | 0.024 | 0.202 | n. a.[d] | −0.43 |
| 275.134 | 2.00743 | 8.050 | 0.024 | 0.295 | 0.05 | −0.13 |
| 275.130 | 2.00241 | 8.030 | 0.024 | 0.296 | 0.05 | −0.14 |
| 275.135 | 2.00205 | 8.025 | 0.024 | 0.296 | 0.01 | −0.17 |
| 275.137 | 1.99959 | 8.016 | 0.024 | 0.297 | 0.02 | −0.16 |
| 275.134 | 1.01290 | 4.063 | 0.023 | 0.574 | 0.20 | 0.11 |
| 275.138 | 1.00639 | 4.040 | 0.023 | 0.577 | 0.28 | 0.20 |
| 275.135 | 1.00229 | 4.023 | 0.023 | 0.580 | 0.27 | 0.18 |
| 275.141 | 1.00204 | 4.020 | 0.023 | 0.580 | 0.22 | 0.13 |
| 275.134 | 1.00193 | 4.023 | 0.023 | 0.580 | 0.29 | 0.21 |
| | | | 300.00 K | | | |
| 300.085 | 4.9429 | 18.021 | 0.023 | 0.636 | −0.15 | −0.61 |
| 300.085 | 3.9963 | 14.595 | 0.023 | 0.478 | −0.11 | −0.47 |
| 300.085 | 3.00062 | 10.974 | 0.024 | 0.194 | −0.07 | −0.34 |
| 300.085 | 2.03943 | 7.466 | 0.025 | 0.153 | −0.05 | −0.22 |
| 300.086 | 1.01782 | 3.730 | 0.024 | 0.379 | 0.02 | −0.06 |
| | | | 325.00 K | | | |
| 325.052 | 7.0028 | 23.302 | 0.026 | 0.110 | −0.27 | −0.83 |
| 325.051 | 6.0009 | 20.004 | 0.025 | 0.126 | −0.36 | −0.84 |
| 325.051 | 5.0023 | 16.699 | 0.025 | 0.148 | −0.46 | −0.86 |
| 325.050 | 4.0029 | 13.377 | 0.024 | 0.182 | −0.59 | −0.90 |
| 325.050 | 2.98937 | 9.998 | 0.024 | 0.240 | −0.70 | −0.93 |
| 325.051 | 2.00149 | 6.706 | 0.024 | 0.352 | −0.69 | −0.84 |
| 325.054 | 1.00177 | 3.374 | 0.023 | 0.689 | −0.33 | −0.40 |
| | | | 350.00 K | | | |
| 350.039 | 7.0060 | 21.532 | 0.025 | 0.159 | −0.37 | −0.85 |
| 350.041 | 6.0048 | 18.500 | 0.024 | 0.196 | −0.44 | −0.86 |
| 350.042 | 5.0041 | 15.449 | 0.024 | 0.259 | −0.53 | −0.88 |
| 350.043 | 4.0017 | 12.376 | 0.024 | 0.379 | −0.63 | −0.91 |
| 350.043 | 2.98518 | 9.246 | 0.023 | 0.743 | −0.75 | −0.96 |
| 350.043 | 2.00226 | 6.216 | 0.023 | 0.444 | −0.76 | −0.90 |
| 350.034 | 1.69695 | 5.286 | 0.026 | 0.085 | −0.49 | −0.60 |
| 350.046 | 1.00200 | 3.126 | 0.023 | 0.743 | −0.49 | −0.55 |
| 350.038 | 1.00122 | 3.124 | 0.026 | 0.094 | −0.49 | −0.56 |
| | | | 375.00 K | | | |
| 375.055 | 7.1165 | 20.398 | 0.025 | 0.123 | −0.09 | −0.54 |
| 375.056 | 6.0007 | 17.265 | 0.025 | 0.144 | −0.10 | −0.48 |
| 375.053 | 5.0035 | 14.444 | 0.025 | 0.170 | −0.10 | −0.42 |
| 375.055 | 4.0028 | 11.593 | 0.024 | 0.209 | −0.09 | −0.35 |
| 375.054 | 2.98467 | 8.672 | 0.024 | 0.275 | −0.08 | −0.27 |
| 375.053 | 2.00265 | 5.835 | 0.024 | 0.403 | −0.09 | −0.22 |
| 375.062 | 1.00311 | 2.930 | 0.023 | 0.792 | −0.11 | −0.17 |

(a) $U(T)$ = 15 mK

(b) $\frac{U(p>3)}{\text{MPa}} = 75 \cdot 10^{-6} \cdot \frac{p}{\text{MPa}} + 3.5 \cdot 10^{-3}$; $\frac{U(p<3)}{\text{MPa}} = 60 \cdot 10^{-6} \cdot \frac{p}{\text{MPa}} + 1.7 \cdot 10^{-4}$

(c) $\frac{U(\rho)}{\text{kg·m}^{-3}} = 2.5 \cdot 10^{4} \frac{\chi_{\text{S}}}{\text{m}^{3}\text{kg}^{-1}} + 1.1 \cdot 10^{-4} \cdot \frac{\rho}{\text{kg·m}^{-3}} + 2.3 \cdot 10^{-2}$

(d) not available, because REFPROP 10.0 does not calculate a converging value of the density at this state point using the AGA8-DC92 EoS [48,49].

**Table 7.** Statistical analysis of the ($p$, $\rho$, $T$) data sets with respect to the AGA8-DC92 EoS [48,49], the GERG-2008 EoS [50,51], the virial EoS [Equation 13], and the PC-SAFT EoS [55–57] for all three ($H_2$ + $C_3H_8$) mixtures studied in this work, including literature data for comparable mixtures. *AARD* = absolute average relative deviation, *Bias* = average deviation, *RMS* = root mean square deviation, Max*RD* = maximum relative deviation.

| Reference[a] | $x(C_3H_8)$ | $N$[b] | Covered ranges | | Experimental *vs* AGA8-DC92 EoS | | | | Experimental *vs* GERG-2008 EoS | | | | Experimental *vs* virial EoS | | Experimental *vs* PC-SAFT EoS | |
|---|---|---|---|---|---|---|---|---|---|---|---|---|---|---|---|---|
| | | | $T$ / K | $p$ / MPa | *AARD* / % | *Bias* / % | *RMS* / % | Max*RD* / % | *AARD* / % | *Bias* / % | *RMS* / % | Max*RD* / % | *RMS* / % | Max*RD* / % | *RMS* / % | Max*RD* / % |
| This work | 0.0499650 | 102 | 250–375 | 1–20 | 0.080 | −0.013 | 0.10 | 0.31 | 0.21 | −0.19 | 0.25 | 0.52 | 0.050 | 0.21 | 0.083 | 0.18 |
| This work | 0.0999278 | 86 | 275–375 | 1–20 | 0.15 | −0.14 | 0.19 | 0.47 | 0.50 | −0.50 | 0.56 | 0.91 | 0.044 | 0.12 | 0.091 | 0.17 |
| This work | 0.1693740 | 39 | 275–375 | 1–7 | 0.28 | −0.22 | 0.36 | 0.76 | 0.54 | −0.50 | 0.62 | 0.96 | — | — | 0.31 | 0.73 |
| Mason and Eakin [30] | 0.4949 | 1 | 288.706 | 0.101325 | 0.0007 | 0.0007 | 0.0007 | 0.0007 | 0.033 | 0.033 | 0.033 | 0.033 | 0.003 | 0.003 | 0.002 | 0.002 |
| Mason and Eakin [30] | 0.4993 | 1 | 288.706 | 0.101325 | 0.004 | −0.004 | 0.004 | 0.004 | 0.029 | 0.029 | 0.029 | 0.029 | 0.008 | 0.008 | 0.007 | 0.007 |
| Mihara et al. [29] [c] | 0.161 | 12 | 298.15 | 0.35–3.8 | 0.039 | −0.039 | 0.047 | 0.089 | 0.17 | −0.17 | 0.21 | 0.43 | 0.14 | 0.31 | 0.067 | 0.15 |
| Mihara et al. [29] [c] | 0.199 | 37 | 323.15–348.15 | 0.31–5.1 | 0.015 | −0.003 | 0.022 | 0.073 | 0.10 | −0.10 | 0.14 | 0.37 | 0.11 | 0.35 | 0.077 | 0.19 |
| Mihara et al. [29] [c] | 0.266 | 24 | 323.15–348.15 | 0.28–3.3 | 0.030 | 0.029 | 0.033 | 0.056 | 0.11 | −0.11 | 0.16 | 0.42 | 0.079 | 0.24 | 0.11 | 0.26 |

[a] Only vapor and supercritical phase measurements have been considered.

[b] Number of experimental points.

[c] Used only for validation in the original GERG-2008 EoS [50,51] work.

**Table 8.** Virial coefficients $B(T)$ and $C(T)$, as well as the second interaction virial coefficients $B_{12}(T)$, with their expanded ($k$ = 2) uncertainties, from the fit to the experimental binary ($H_2$ (1) + $C_3H_8$ (2)) mixtures studied in this work at the average temperature of each isotherm. Relative deviations from the coefficients given by the AGA8-DC92 EoS [48,49], $B_{12,AGA8-DC92}$, and the GERG-2008 EoS [50,51], $B_{12,GERG-2008}$.

| $T$ / K | $B$ / $cm^3 \cdot mol^{-1}$ | $U(B)$ / $cm^3 \cdot mol^{-1}$ | $C$ / $cm^6 \cdot mol^{-2}$ | $U(C)$ / $cm^6 \cdot mol^{-2}$ | $B_{12}$ / $cm^3 \cdot mol^{-1}$ | $U(B_{12})$ / $cm^3 \cdot mol^{-1}$ | $B_{12,exp} - B_{12,AGA8-DC92}$ / $cm^3 \cdot mol^{-1}$ | $B_{12,exp} - B_{12,GERG-2008}$ / $cm^3 \cdot mol^{-1}$ |
|---|---|---|---|---|---|---|---|---|
| | | | | (0.95 $H_2$ + 0.05 $C_3H_8$) | | | | |
| 275.109 | 11.05 | 0.17 | 570 | 27 | –4.79 | 0.72 | –3.5 | 8.0 |
| 300.047 | 12.15 | 0.20 | 545 | 34 | –0.54 | 0.08 | –3.2 | 5.6 |
| 325.067 | 13.28 | 0.23 | 511 | 42 | 5.69 | 0.81 | –0.9 | 6.2 |
| 350.067 | 13.98 | 0.25 | 501 | 50 | 8.55 | 1.20 | –2.2 | 4.1 |
| 375.060 | 14.37 | 0.26 | 519 | 51 | 9.05 | 1.27 | –6.2 | 0.3 |
| | | | | (0.90 $H_2$ + 0.10 $C_3H_8$) | | | | |
| 300.071 | 8.43 | 0.16 | 730 | 28 | 2.57 | 0.37 | –0.23 | 8.4 |
| 325.079 | 10.85 | 0.14 | 634 | 25 | 10.74 | 1.52 | 3.83 | 11.6 |
| 350.077 | 11.68 | 0.75 | 608 | **237** | 11.13 | 1.70 | 0.38 | 7.5 |
| 375.068 | 12.89 | 0.18 | 618 | 36 | 14.47 | 2.02 | –0.02 | 6.9 |

**Table 9.** Parameters of the interpolation of the second interaction virial coefficient $B_{12}(T)$ for the binary ($H_2$ (1) + $C_3H_8$ (2)) system as a function of temperature using Equation 16.

| Parameter | Value ± Standard Uncertainty | Unit |
|---|---|---|
| $N_0$ | 62 ± 13 | $cm^3 \cdot mol^{-1}$ |
| $N_1$ | –18700 ± 3900 | $cm^3 \cdot mol^{-1} \cdot K$ |
| *RMS* of residuals | 3.0 | $cm^3 \cdot mol^{-1}$ |

**Table 10.** Pure component $m$, $\sigma$, $\varepsilon/k_B$ and binary interaction $k_{12}$ parameters of the PC-SAFT EoS [55–57] fit to the experimental binary ($H_2$ (1) + $C_3H_8$ (2)) mixtures studied in this work.

| Component | $m^{(a)}$ | $\sigma$ / Å$^{(a)}$ | $\varepsilon/k_B$ / K$^{(a)}$ | $k_{12}^{(b)}$ |
| --- | --- | --- | --- | --- |
| Hydrogen (normal) | 0.94 | 2.91 | 25.6 | 0.0927 |
| Propane | 1.629 | 3.867 | 230.9 | |

(a) Pure component parameters of hydrogen (normal) and propane taken from literature [108].

(b) Temperature-dependent binary interaction $k_{12}$ parameter ($k_{12} = A/T + B$): $A$ = 34.9 K, $B$ = −0.0114.

## Figures

**Figure 1.** *p*, *T*-phase diagram showing the experimental points (●) and the calculated phase envelope (solid line) using the GERG-2008 EoS [50,51] for: a) (0.95 $H_2$ + 0.05 $C_3H_8$), b) (0.90 $H_2$ + 0.10 $C_3H_8$), and c) (0.83 $H_2$ + 0.17 $C_3H_8$) binary mixtures, respectively. The marked temperature and pressure ranges represent the range of the binary experimental data used for the development of the GERG-2008 EoS [50,51] (red dashed line), and the area of interest for the hydrogen-based economy systems (black dashed line).

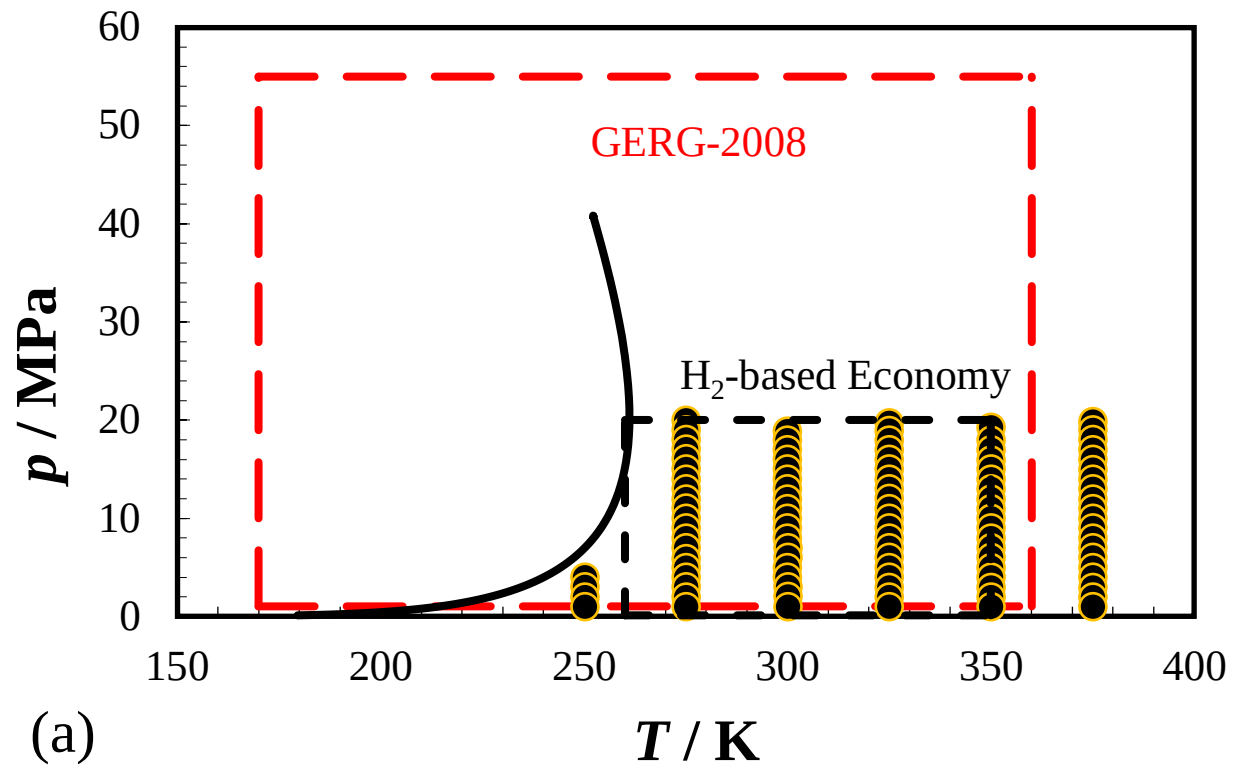

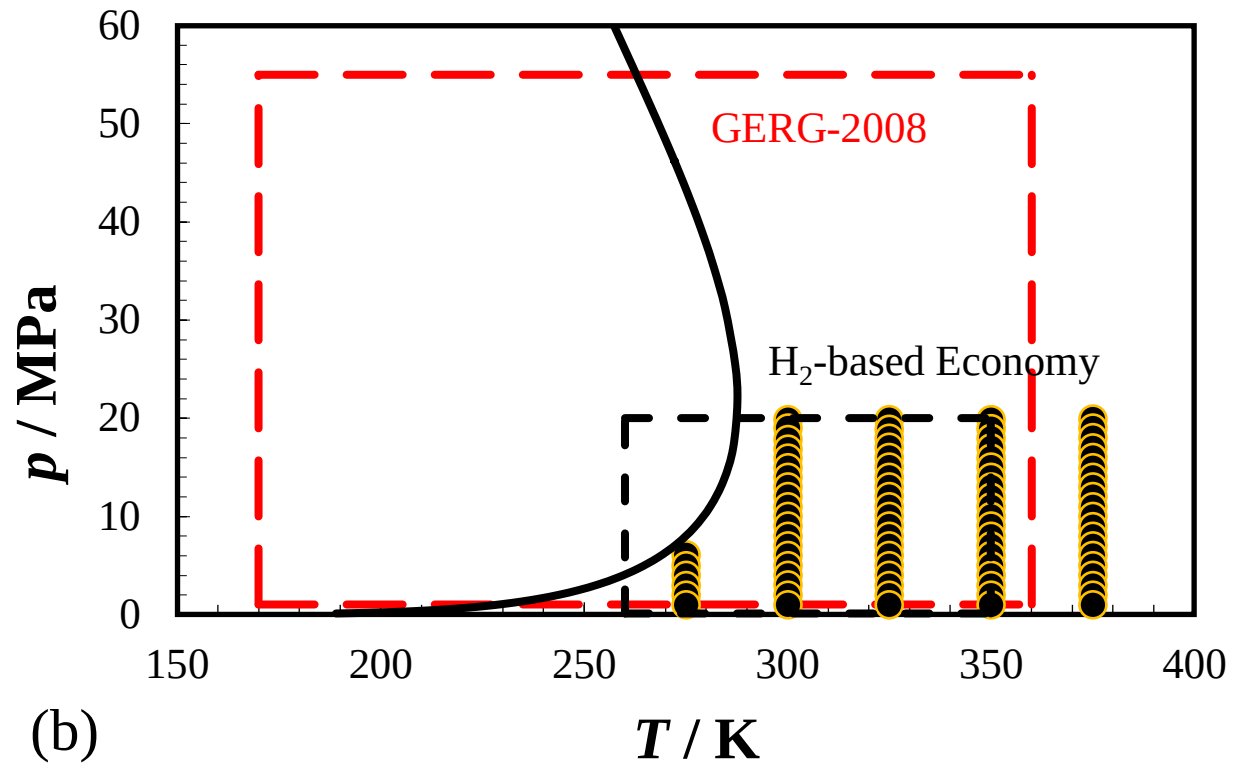

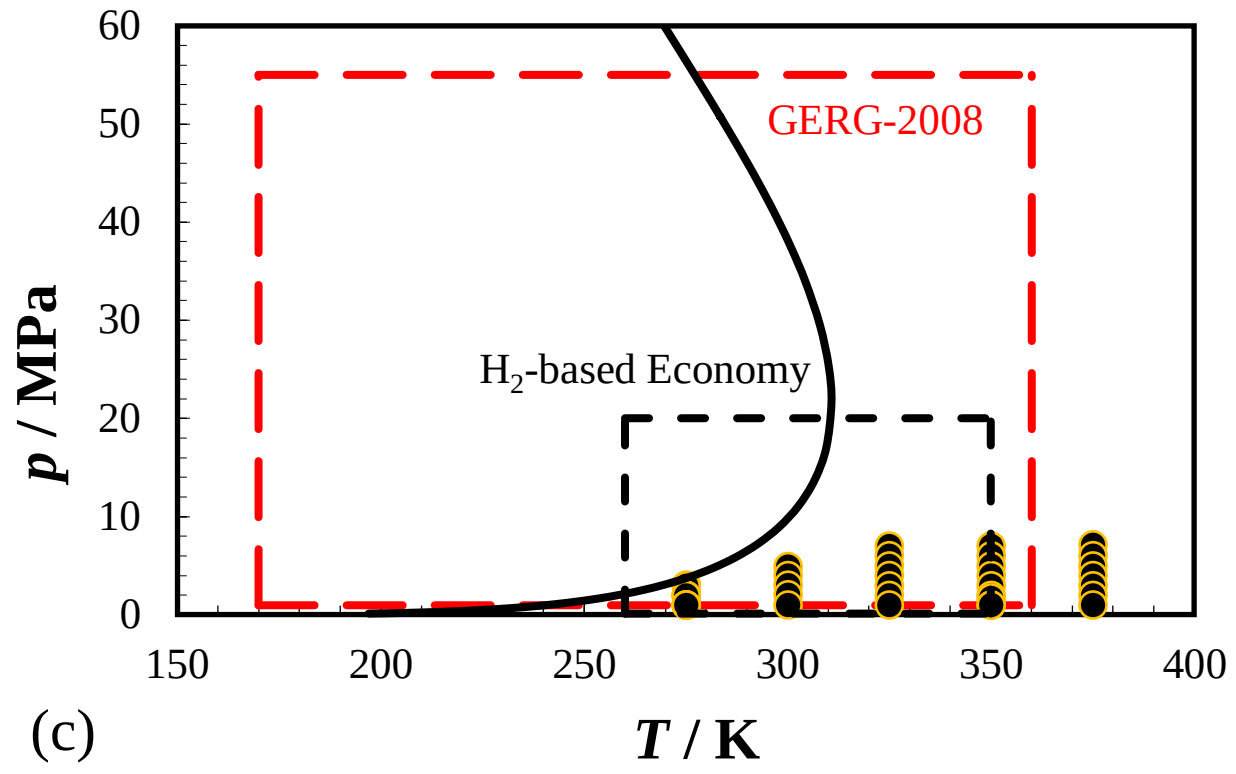

**Figure 2.** Relative deviations of experimental density , $\rho_{exp}$, data of the binary (0.95 $H_2$ + 0.05 $C_3H_8$) mixture from density values calculated from the: (a) AGA8-DC92 EoS [48,49], $\rho_{AGA8\text{-}DC92}$, and (b) GERG-2008 EoS [50,51], $\rho_{GERG\text{-}2008}$, as a function of pressure for different temperatures: □ 250 K, ◇ 275 K, △ 300 K, × 325 K, + 350 K, ○ 375 K. Dashed lines indicate the expanded ($k$ = 2) uncertainty of the corresponding EoS. Error bars exemplarily given on the 325-K data set indicate the expanded ($k$ = 2) uncertainty of the experimental density.

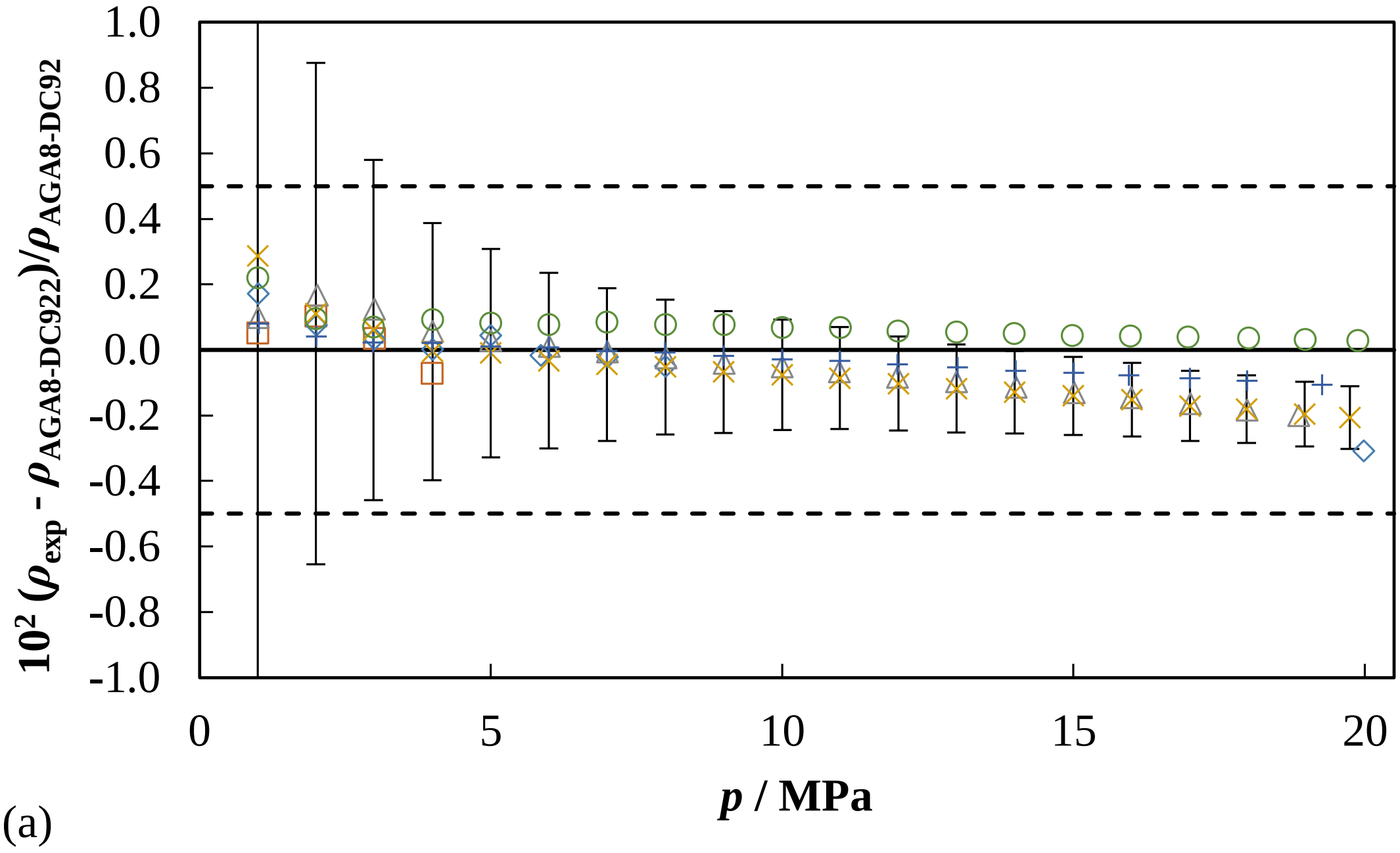

(a)

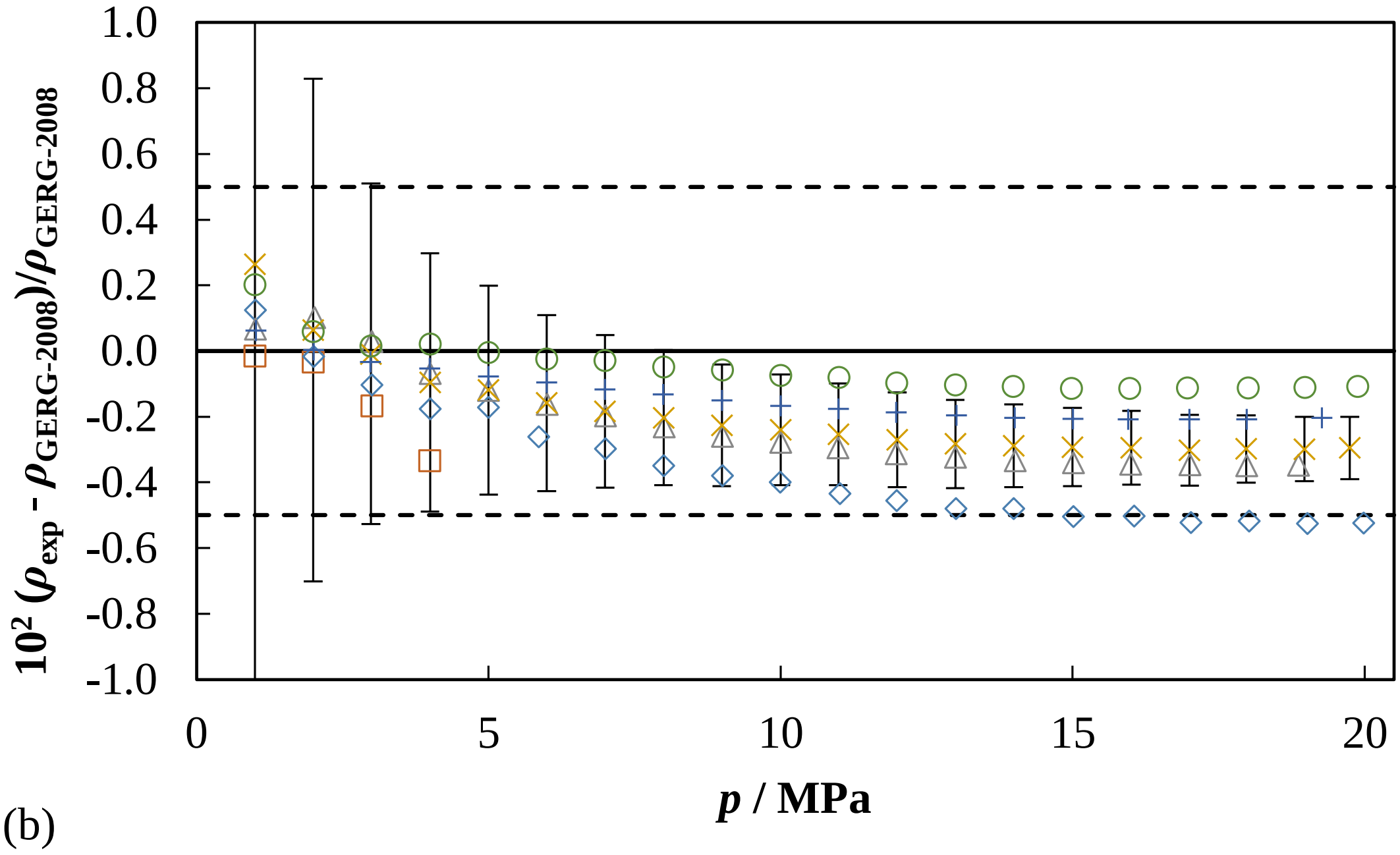

(b)

**Figure 3.** Relative deviations of experimental density, $\rho_{exp}$, data of the binary (0.90 $H_2$ + 0.10 $C_3H_8$) mixture from density values calculated from the: (a) AGA8-DC92 EoS [48,49], $\rho_{AGA8\text{-}DC92}$, and (b) GERG-2008 EoS u [50,51], $\rho_{GERG\text{-}2008}$, as a function of pressure for different temperatures: ◇ 275 K, △ 300 K, × 325 K, + 350 K, ○ 375 K. Dashed lines indicate the expanded ($k$ = 2) uncertainty of the corresponding EoS. Error bars exemplarily given on the 325-K data set indicate the expanded ($k$ = 2) uncertainty of the experimental density.

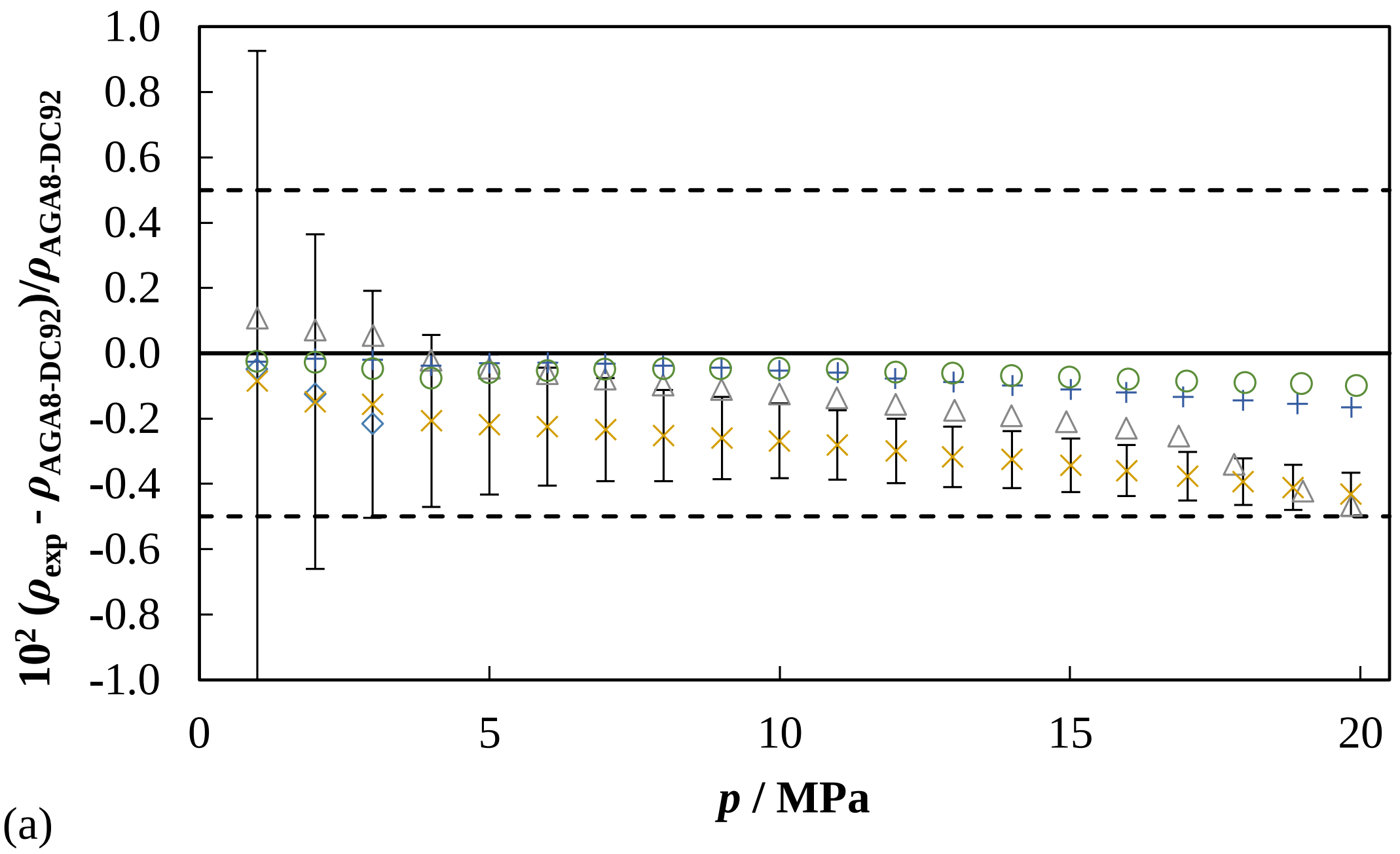

(a)

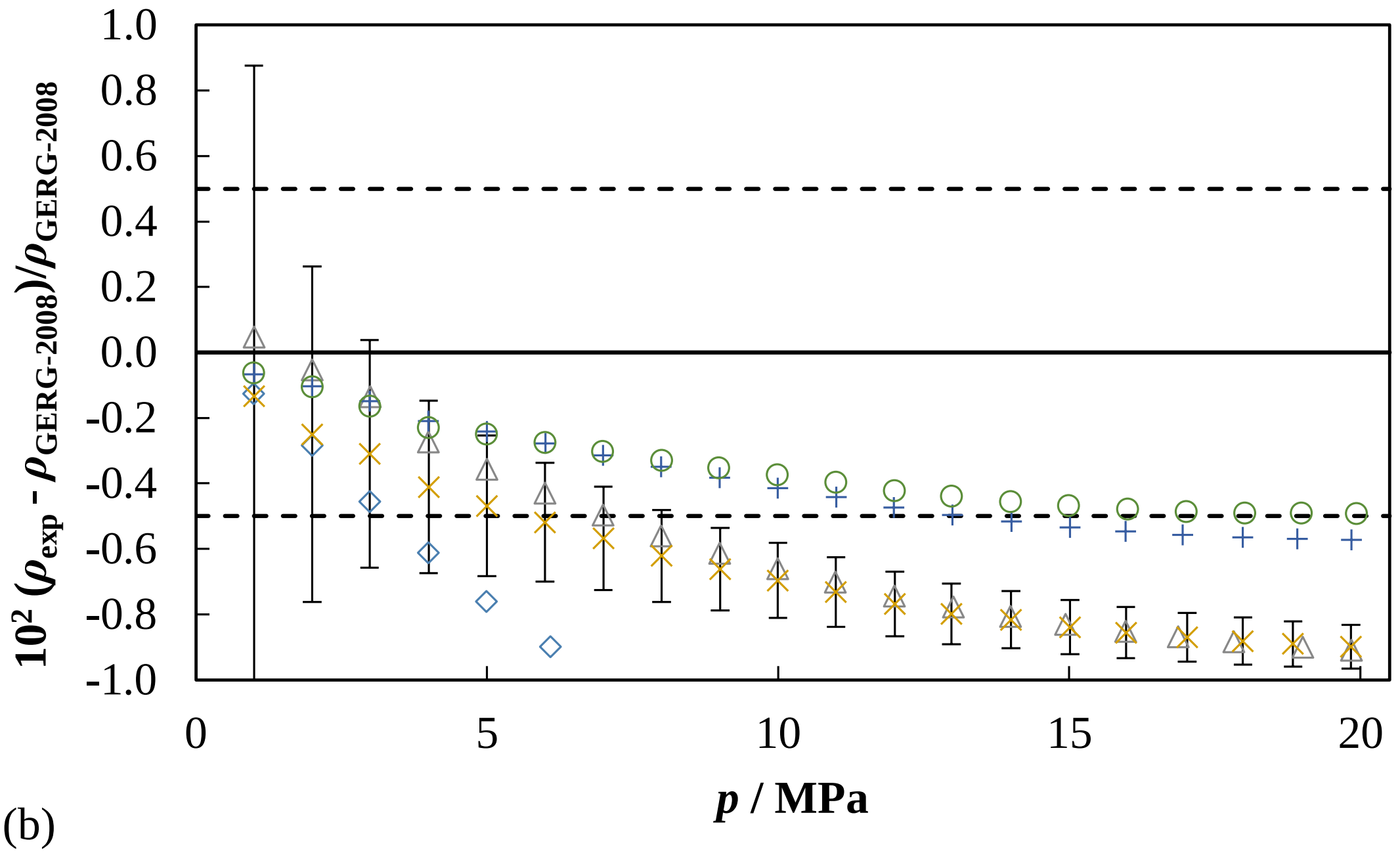

(b)

**Figure 4.** Relative deviations of experimental density , $\rho_{exp}$, data of the binary (0.83 $H_2$ + 0.17 $C_3H_8$) mixture from density values calculated from the: (a) AGA8-DC92 EoS [48,49], $\rho_{AGA8\text{-}DC92}$, and (b) GERG-2008 EoS [50,51], $\rho_{GERG\text{-}2008}$, as a function of pressure for different temperatures: ◇ 275 K, △ 300 K, × 325 K, + 350 K, ○ 375 K. Dashed lines indicate the expanded ($k$ = 2) uncertainty of the corresponding EoS. Error bars exemplarily given on the 325-K data set indicate the expanded ($k$ = 2) uncertainty of the experimental density.

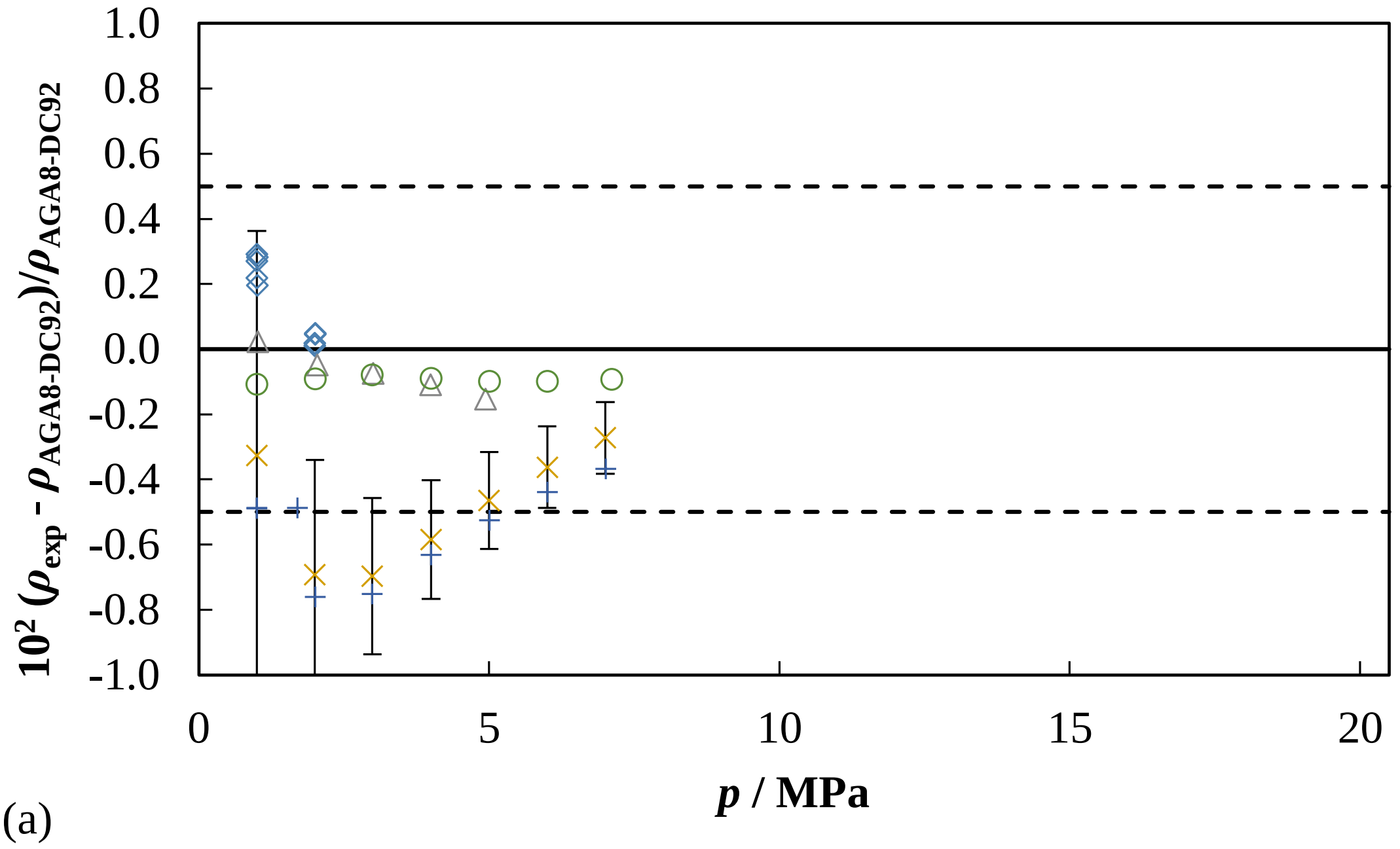

(a)

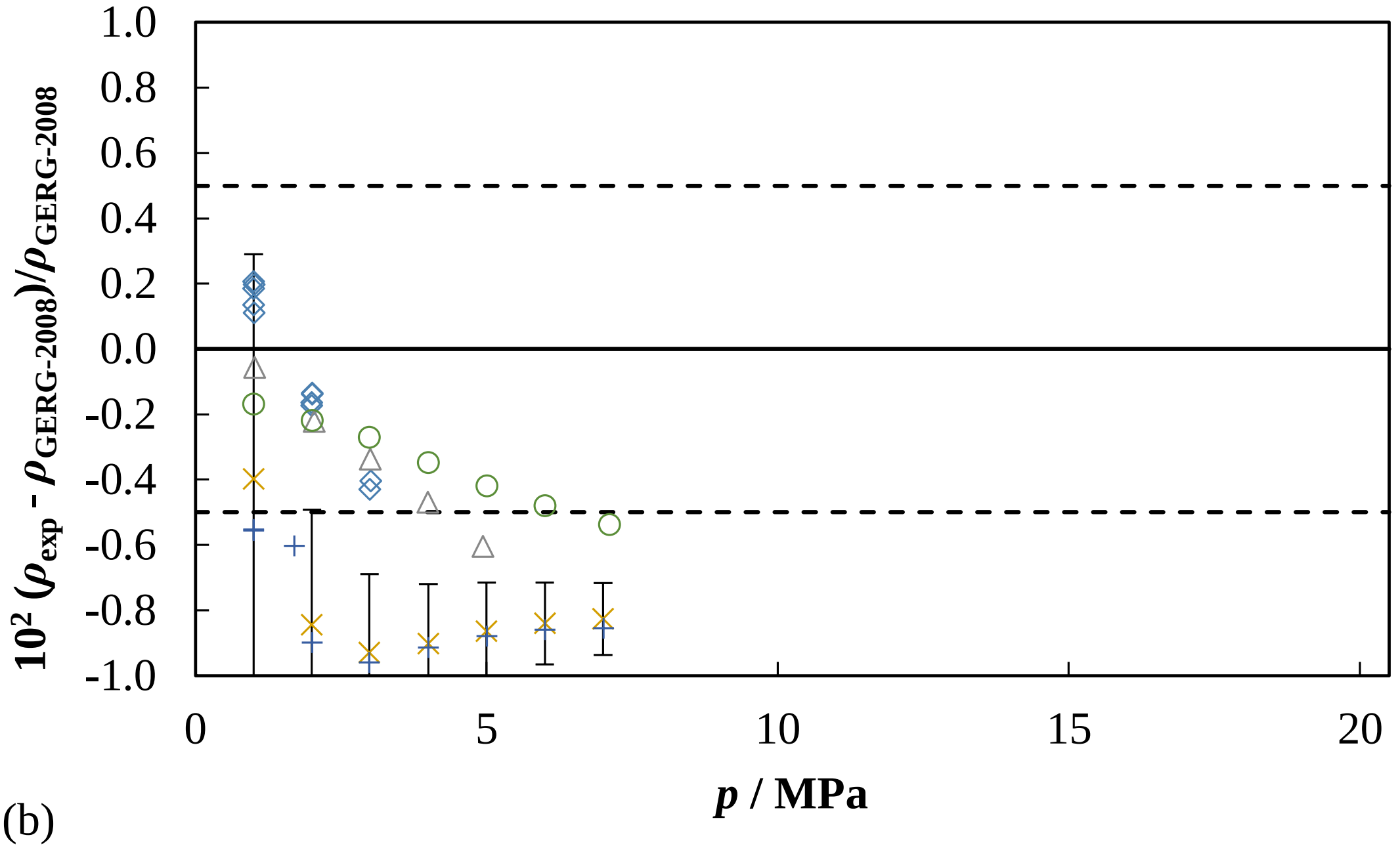

(b)

**Figure 5.** Second interaction virial coefficient $B_{12}(T)$ for the binary ($H_2$ + $C_3H_8$) system estimated from the experimental data as a function of the mole fraction of $C_3H_8$, $x_{C_3H_8}$, at temperatures: ◇ 275 K, △ 300 K, × 325 K, + 350 K, ○ 375 K. The dashed lines represent the $B_{12}(T)$ values computed from the GERG-2008 EoS using the reference pure-fluid equations of state [50,51] at the corresponding temperatures.

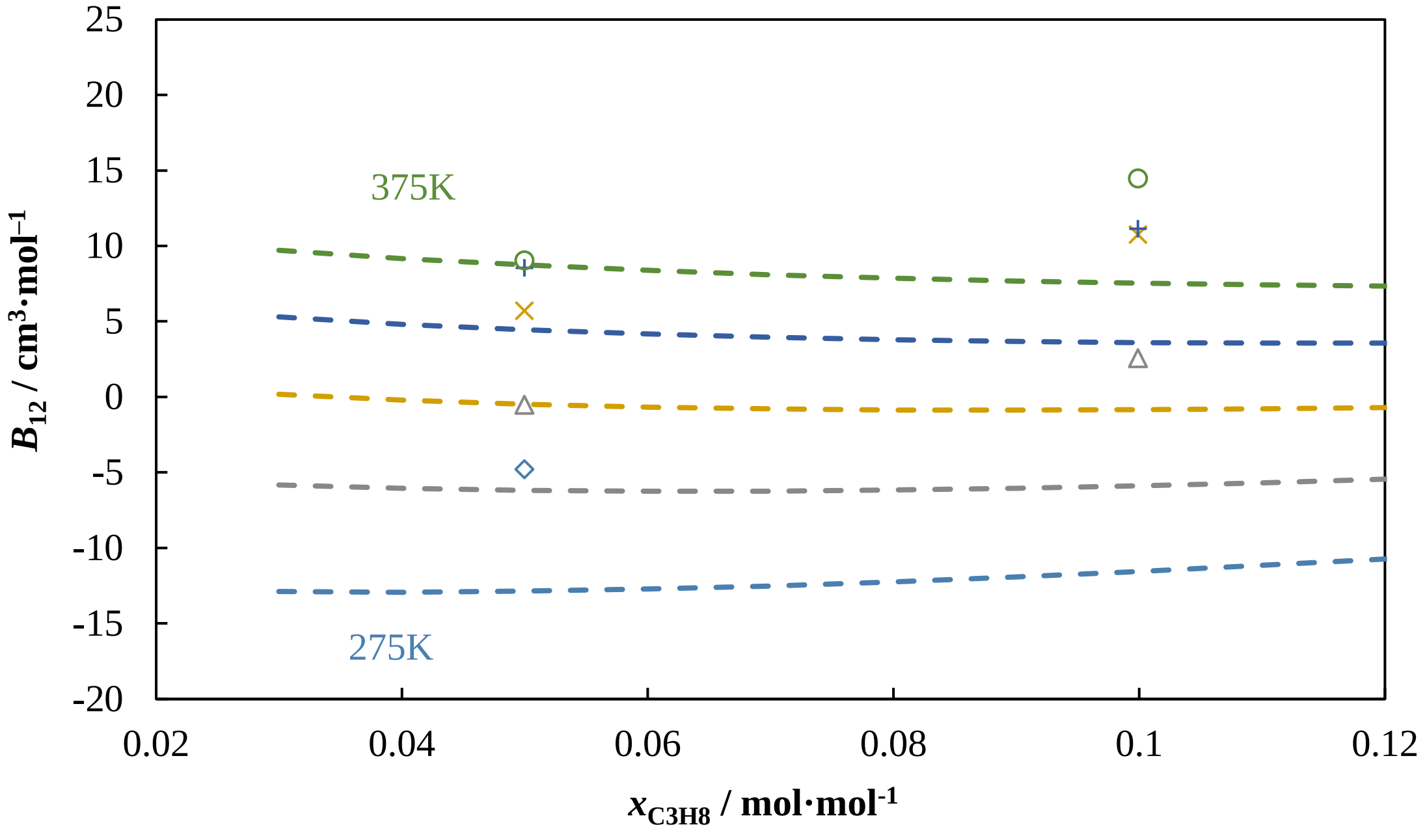

**Figure 6.** (a) Averaged second interaction virial coefficient $B_{12}(T)$ for the binary ($H_2$ + $C_3H_8$) system as a function of temperature from: × this work, □ Brewer [103], ◇ Malesinska et al. [104], △ Mason and Eakin [53], ○ Mihara et al. [52], - - AGA8-DC92 EoS [48,49], ⋯ GERG-2008 EoS [50,51]. Error bars indicate the expanded ($k$ = 2) uncertainty of the estimated $B_{12}(T)$ values, the solid line represents the fit of the experimental data of this work to Equation 16; (b) Residuals of the fit to Equation 16.

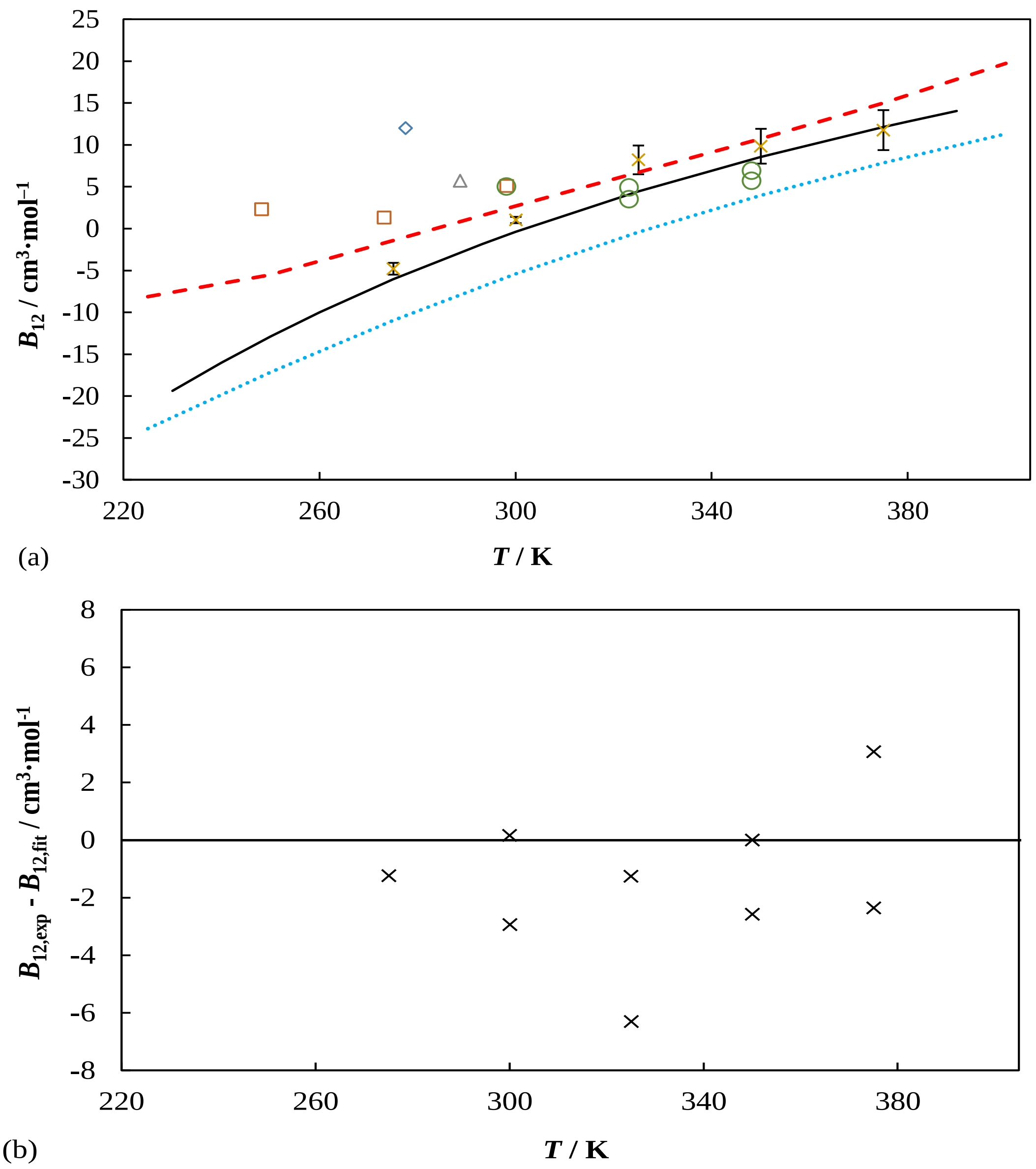

**Figure 7.** Residual plots of experimental ($p$, $\rho_{exp}$, $T$) data of all the binary ($H_2$ + $C_3H_8$) mixtures from the density values calculated from the fitted PC-SAFT EoS [55–57] as a function of density for the different mole fractions of $C_3H_8$, $x_{C_3H_8}$, of this work: □ 0.05, ◇ 0.10, △ 0.17.

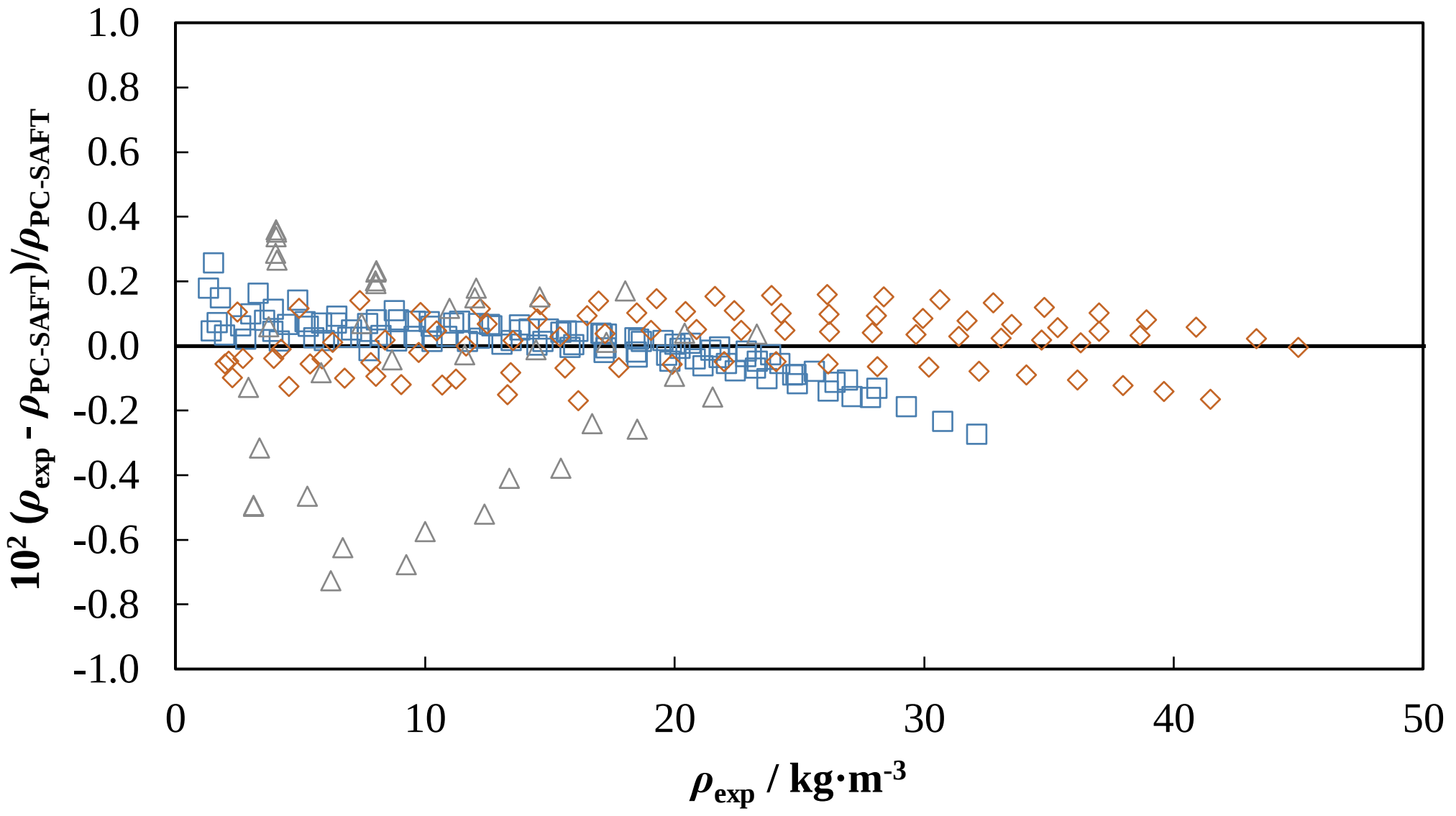